\documentclass[12pt,a4j]{article}
\usepackage[dvips]{graphicx}
\usepackage{amsmath}
\usepackage{amssymb}
\usepackage{enumerate}

\begin{document}
\baselineskip=7mm
\centerline{\bf  A direct method for solving the generalized sine-Gordon equation II}\par
\bigskip
\centerline{Yoshimasa Matsuno}\par
\centerline{ Division of Applied Mathematical Science, Graduate School of Science and Engineering} \par
\centerline{ Yamaguchi University, Ube, Yamaguchi 755-8611, Japan} \par
\centerline{ E-mail: matsuno@yamaguchi-u.ac.jp} \par
\bigskip
\bigskip
\noindent {\bf Abstract}\par
\noindent  The generalized sine-Gordon (sG) equation $u_{tx}=(1+\nu\partial_x^2)\sin\,u$ was derived as an integrable
generalization of the sG equation. In a previous paper (Matsuno Y 2010 J. Phys. A: Math. Theor. {\bf 43}  105204) which is
referred to as I, we developed
a systematic method for solving the generalized sG equation with $\nu=-1$. Here, we address the equation with $\nu=1$.
By solving the equation analytically, we find that the structure of solutions differs substantially from that of the
former equation.
In particular, we show that the equation exhibits kink and breather solutions and does not admit multi-valued solutions like loop
solitons as obtained in I. We also demonstrate that the equation reduces to the short pulse and sG equations in
appropriate scaling limits. 
The limiting forms of the multisoliton solutions are also presented.
Last, we provide a recipe for deriving an infinite number of conservation laws
by using a novel B\"acklund transformation connecting solutions of the sG and generalized sG equations. 
 \par
\noindent Mathematics Classification: 35Q51, 37K10, 37K40 \par

\bigskip
\bigskip
\newpage
\leftline{\bf  1. Introduction} \par
\bigskip
\noindent  The generalized sine-Gordon (sG) equation
$$u_{tx}=(1+\nu\partial_x^2)\sin\,u, \eqno(1.1)$$
where $u=u(x,t)$ is a scalar-valued function, $\nu$ is a real parameter,  $\partial_x^2=\partial^2/\partial x^2$ 
and the subscripts $t$ and $x$ appended to $u$ denote partial differentiation,
has been derived  in [1] using bi-Hamiltonian methods.  In the case of $\nu=-1$, its
integrability was established by constructing a Lax pair associated with it  and the initial
value problem was formulated for decaying initial data by means of the inverse scattering method [2].
Quite recently, we developed a systematic method for solving equation (1.1) with
$\nu=-1$ and obtained soliton solutions in the form of parametric representation [3],
which will be referred to as I hereafter. We showed that
the solutions exhibit various new features which have never seen in the sG solitons.
 We also demonstrated that the
generalized sG equation reduces to the physically important short pulse and sG equations in appropriate scaling limits. \par
In this paper, we consider equation (1.1) with $\nu=1$
$$u_{tx}=(1+\partial_x^2)\sin\,u. \eqno(1.2)$$
Despite its resemblance to equation (1.1) with $\nu=-1$, the analysis of equation (1.2) by the inverse scattering method has not been done and explicit solutions are
still unavailable [2].
This is a motivation why we address equation (1.2).  We use an exact method of solution  developed in I to construct soliton solutions
and investigate their properties. 
One of the remarkable features of equation (1.2) is that it does not admit multi-valued solutions like loop solitons as obtained in I.
It is therefore of considerable interest to study the structure of solutions in conparison with those of equation (1.1) with $\nu=-1$.
\par
This paper is organized as follows. In section 2, we summarize a direct method of solution  
in such a manner that is relevant to equation (1.2). 
Specifically, a hodograph transformation makes it possible to transform equation (1.2) into an integrable  system of nonlinear partial
differential equations (PDEs). This system is further recast to a system of bilinear equations through the dependent variable transformations.
The solutions to the bilinear equations can be constructed following the standard procedure in the bilinear transformation method [4, 5].
Inverting the relationship that determines the hodograph transformation, we obtain a parametric representation of the solution of equation (1.2)
in terms of tau functions.
In section 3, we present soliton solutions and investigate their properties focusing mainly on 1- and 2-soliton and general multisoliton
solutions. 
Throughout this paper, we use the term 'soliton' as a generic name of elementary solutions such as kink, antikink and breather solutions.
In section 4, we show that equation (1.2) reduces to the short pulse equation in an appropriate scaling limit. 
The limiting form of the multisoliton solution is also presented.
We find that the short pulse equation exhibits a novel type of singular soliton solutions.
Last, we briefly
discuss the reduction to the sG equation, reproducing the known results about the multisoliton solution and its characteristics.
In section 5, we provide a recipe for deriving an infinite number of conservation laws starting from those of the sG equation. It is based
on a novel B\"acklund transformation connecting solutions of the sG and generalized sG equations.
 Section 6 is devoted to conclusion where we discuss some open problems associated with the generalized sG equation. 
In Appendix A, we present a method for obtaining solutions of traveling type  and recover the 1-soliton solutions.
\par
 \par
 \bigskip
 \leftline{\bf 2. Exact method of solution} \par
 \bigskip
 \leftline{\it 2.1. Hodograph transformation} \par
 \medskip
 \noindent 
 Here, we summarize an exact method of solution which consists of a sequence of nonlinear transformations.
 First, we introduce  the new dependent variable $r$ in accordance with the relation
$$r^2=1-u_x^2,\qquad (0<r<1), \eqno(2.1)$$
to transform equation (1.2)  into the conservation law of the form
$$ r_t-(r\,\cos\,u)_x=0. \eqno(2.2)$$
This expression makes it possible to define the hodograph transformation $(x ,t) \rightarrow (y, \tau)$ by
$$dy=rdx+r\,\cos\,u\,dt, \qquad d\tau=dt. \eqno(2.3)$$ 
 The $x$ and $t$ derivatives are then rewritten  in terms of the $y$ and $\tau$ derivatives  as
 $${\partial\over\partial x}=r{\partial\over\partial y}, \qquad {\partial\over\partial t}
 ={\partial\over\partial \tau}+r\,\cos\,u\,{\partial\over\partial y}. \eqno(2.4)$$
 With the new variables $y$ and $\tau$, (2.1) and (2.2) are recast into the form
 $$r^2=1-r^2u_y^2, \eqno(2.5)$$
 $$\left({1\over r}\right)_\tau+(\cos\,u)_y=0, \eqno(2.6)$$
 respectively. Further reduction is possible if one defines the variable $\phi$ by
 $$u_y=\sinh\,\phi, \qquad \phi=\phi(y,\tau). \eqno(2.7)$$
  It follows from (2.5) and (2.7) that
  $$ {1\over r}=\cosh\,\phi. \eqno(2.8)$$
 Substituting (2.7) and (2.8) into equation (2.6), we find
 $$\phi_\tau=\sin\,u. \eqno(2.9)$$ 
 If we eliminate the variable $\phi$ from (2.7) and (2.9), we obtain a single PDE for $u$
 $${u_{\tau y}\over \sqrt{1+u_y^2}}=\sin\,u. \eqno(2.10)$$
 Similarly, elimination of the variable $u$ gives a single PDE for $\phi$
 $${\phi_{\tau y}\over \sqrt{1-\phi_\tau^2}}=\sinh\,\phi. \eqno(2.11)$$
 \par
 By inverting the hodograph transformation (2.4) and using (2.8), the equation that determines the inverse mapping $(y,\tau) \rightarrow (x,t)$ is  found to be governed 
by the system of linear  PDEs for $x=x(y,\tau)$
$$x_y=\cosh\,\phi, \eqno(2.12a)$$
$$x_\tau=-\cos\,u. \eqno(2.12b)$$ 
 It is important that the integrability of the system of equations (2.12) is assured by (2.7) and (2.9). 
 Thus, given $\phi$ and $u$, the above system of equations can be integrated to give the parametric expression of $x$
 in terms of $y$ and $\tau$.  Remarkably, we were able to perform the integration analytically. 
 This last step for constructing solutions   is the core of the present analysis.
 The main results associated with soliton solutions will be given by theorem 2.1 and theorem 2.2 below.
 \par
 \bigskip
 \leftline{\it 2.2. Bilinear formalism}\par
 \medskip
 \noindent Here, we develop a method for solving a system of PDEs (2.7) and (2.9). 
 Let $\sigma$ and $\sigma^\prime$
 be solutions of the sG equation
 $$\sigma_{\tau y}=\sin\,\sigma, \qquad \sigma=\sigma(y,\tau),\eqno(2.13a)$$
 $$\sigma^\prime_{\tau y}=\sin\,\sigma^\prime, \qquad \sigma^\prime=\sigma^\prime(y,\tau). \eqno(2.13b)$$
 The  solutions of the above equations can be put into the form [6-9]
 $$\sigma=2{\rm i}\,\ln{f^\prime\over f}, \eqno(2.14a)$$
 $$\sigma^{\prime}=2{\rm i}\, \ln{g^\prime\over g}. \eqno(2.14b)$$
 For soliton solutions, the tau functions $f, f^\prime, g$ and $g^\prime$ satisfy the following system of bilinear equations [6]:
 $$D_\tau D_yf\cdot f={1\over 2}(f^2-{f^\prime}^2), \eqno(2.15a)$$
 $$D_\tau D_yf^\prime\cdot f^\prime={1\over 2}({f^\prime}^2-f^2), \eqno(2.15b)$$
  $$D_\tau D_yg\cdot g={1\over 2}(g^2-{g^\prime}^2), \eqno(2.16a)$$
 $$D_\tau D_yg^\prime\cdot g^\prime={1\over 2}({g^\prime}^2-g
 ^2), \eqno(2.16b)$$
  where the bilinear operators $D_\tau$ and $D_y$ are defined by
 $$D_\tau^mD_y^nf\cdot g=\left(\partial_\tau-\partial_{\tau^\prime} \right)^m
                 \left(\partial_y-\partial_{y^\prime} \right)^nf(\tau,y)g(\tau^\prime,y^\prime)|_{\tau^\prime=\tau,\,
                 y^\prime=y}, \qquad (m, n=0, 1, 2, ...).   \eqno(2.17)$$
                 \par
 Now, we seek solutions of equations (2.7) and (2.9) of the form 
 $$u={\rm i}\,\ln{F^\prime\over F}, \eqno(2.18a)$$
 $$\phi=\ln{G^\prime\over G}, \eqno(2.18b)$$
 where  $F, F^\prime, G$ and $G^\prime$ are new tau functions. If we impose the condition 
 $$F^\prime F=G^\prime G, \eqno(2.19)$$
 among these tau functions, then  equations (2.7) and (2.9) can be transformed to the following bilinear equations
$${\rm i}\,D_yF^\prime\cdot F={1\over 2}({G^\prime}^2-G^2), \eqno(2.20)$$
 $${\rm i}\,D_\tau G^\prime\cdot G={1\over 2}(F^2-{F^\prime}^2), \eqno(2.21)$$
  respectively.
   The  proposition below provides the tau functions $F, F^\prime, G$ and $G^\prime$ in terms of $f, f^\prime, g$ and $g^\prime$. \par
  \medskip
  \noindent {\bf Proposition 2.1.}\ {\it If we impose the conditions for the tau functions $f, f^\prime, g$ and $g^\prime$
  $${\rm i}\, D_yf\cdot g^\prime={1\over 2}(fg^\prime-f^\prime g), \eqno(2.22a)$$               
  $${\rm i}\, D_y f^\prime\cdot g={1\over 2}(f^\prime g-fg^\prime ), \eqno(2.22b)$$
  $${\rm i}\, D_\tau f\cdot g=-{1\over 2}(fg-f^\prime g^\prime), \eqno(2.23a)$$               
  $${\rm i}\, D_\tau f^\prime\cdot g^\prime=-{1\over 2}(f^\prime g^\prime-fg), \eqno(2.23b)$$
 then the solutions of bilinear equations (2.20) and (2.21) subjected to the condition (2.19) are given by} 
  $$F=fg, \quad F^\prime=f^\prime g^\prime,\eqno(2.24a)$$
  $$ \quad G=fg^\prime, \quad G^\prime=f^\prime g. \eqno(2.24b)$$
    \medskip
  \noindent{\bf Proof.}\ It is obvious that the tau functions (2.24) satisfy the condition (2.19).  We first  prove 
  (2.20). Substituting (2.24) into the left-hand side of (2.20) and using (2.22), we find that
    \begin{alignat*}{1}
  {\rm i}\,D_yF^\prime\cdot F &={\rm i}\{(D_yf^\prime\cdot g)fg^\prime-(D_yf\cdot g^\prime)f^\prime g\} \\
             &=-{1\over 2}(fg^\prime)^2+{1\over 2}(f^\prime g)^2 \\
             &={1\over 2}({G^\prime}^2-G^2), 
  \end{alignat*}
  which is (2.20).   The proof of (2.21) can be done in the same way by using (2.23). 
 \hspace{\fill} $\square$ \par
    \bigskip
  \leftline{\it 2.3. Parametric representation}\par
  \medskip
  \noindent We demonstrate that the solution of equation (1.2) admits a parametric representation.
  The following relation is crucial to integrate (2.12):\par
  \medskip
  \noindent{\bf Proposition 2.2.} {\it $\cosh\,\phi$ is given in terms of the tau functions $f, f^\prime, g$ and $g^\prime$ as}
  $$\cosh\,\phi=1+{\rm i}\,\left(\ln{g^\prime g\over f^\prime f}\right)_y. \eqno(2.25)$$
  \medskip
  \noindent {\bf Proof.} Using (2.22), one obtains
  \begin{alignat*}{1}
   {\rm i}\, \left(\ln{g^\prime g\over f^\prime f}\right)_y &=-{\rm i}\,{D_yf\cdot g^\prime \over fg^\prime }-{\rm i}\,{D_yf^\prime \cdot g\over f^\prime g} \\
                                      &={1\over 2}{(fg^\prime)^2+(f^\prime g)^2 \over f^\prime fg^\prime g}-1. \tag{2.26}
  \end{alignat*}
  On the other hand, it follows from (2.18b) and (2.24b) that
  \begin{alignat*}{1}
  \cosh\,\phi &={1\over 2}\left({G^\prime\over G}+{G\over G^\prime}  \right) \\
             &= {1\over 2}{(fg^\prime)^2+(f^\prime g)^2\over f^\prime fg^\prime g}. \tag{2.27}
  \end{alignat*}
  The relation (2.25)  follows immediately by comparing (2.26) and (2.27). \hspace{\fill}$\square$ \par
  Integrating  (2.12a) with (2.25) by $y$  yields the expression of $x$
  $$x=y+ {\rm i}\,\ln{g^\prime g\over f^\prime f}+d(\tau), \eqno(2.28)$$
  where $d$ is an integration constant which depends generally on $\tau$. The expression (2.28)
  now leads to our main result: \par
  \medskip
  \noindent{\bf Theorem 2.1.} {\it The  solution of equation (1.2)  can be expressed by the
  parametric representation 
   $$u(y,\tau)={\rm i}\,\ln{f^\prime g^\prime \over fg}, \eqno(2.29a)$$
  $$x(y,\tau)=y-\tau + {\rm i}\,\ln{g^\prime g\over f^\prime f} + y_0, \eqno(2.29b)$$
  where the tau functions $f, f^\prime, g$ and $g^\prime$  satisfy equations (2.15), (2.16), (2.22) and (2.23)
  and $y_0$ is an arbitrary constant independent of $y$ and $\tau$}. \par
  \medskip
  \noindent{\bf Proof.} The expression (2.29a) for $u$ is a consequence of (2.18a) and (2.24a).
  To prove (2.29b), we substitute (2.28) into (2.12b) and obtain the relation
  $${\rm i}\,\left(\ln{g^\prime g\over f^\prime f}\right)_\tau+d^\prime(\tau) = -\cos\,u. \eqno(2.30)$$
  The left-hand side of (2.30) is modified by using (2.23) whereas
  the right-hand side  can be expressed by $f, f^\prime, g$ and $g^\prime$ in view of (2.29a).
   After a few calculations, we find that
  most terms are cancelled, leaving  
  the equation $ d^\prime(\tau)=-1$. Integrating this equation, one obtains   $d(\tau)=-\tau+y_0$,
  which, substituted into (2.28),  gives the expression (2.29b) for $x$. \hspace{\fill}$\square$ \par
  \medskip
    An interesting feature of
    the parametric solution (2.29) is that it never exhibits
    singularities as encountered in the case of equation
    (1.1) with $\nu=-1$ (see I).
    To demonstrate  this, we calculate $u_x$ from (2.7) and (2.8) and obtain
  $$u_x=ru_y=\tanh\,\phi. \eqno(2.31)$$
  The relation (2.31) implies that $u_x$ always takes  finite value and singular solutions such as loop solitons and multi-valued kinks
   obtained in I never exist for equation (1.2). See Figures 2-4 in I.
  It should be remarked, however, that the above relation does not exclude the presence of singular solutions like peaked waves (or peakons),
  for instance  which
  have a finite discontinuity in their slope at the crest. The construction of the latter type of solutions is of some interest from the mathematical
  point of view.
  \par
    \bigskip
      \leftline{\it 2.4. Multisoliton solutions} \par
  \medskip
  \noindent The last step in constructing solutions is to find the tau-functions $f, f^\prime, g$ and $g^\prime$ for the
  sG equation which satisfy simultaneously the bilinear equations (2.22) and (2.23).  The following theorem
  establishes this purpose: \par
  \medskip
  \noindent{\bf Theorem 2.2.} {\it The tau-functions $f, f^\prime, g$ and $g^\prime$ given below satisfy both the bilinear forms (2.15) and (2.16) of the
  gG equation  and the
  bilinear equations (2.22) and (2.23),
  $$f=\sum_{\mu=0,1}{\rm exp}\left[\sum_{j=1}^N\mu_j\left(\xi_j+d_j+{\pi\over 2}\,{\rm i}\right)
+\sum_{1\le j<k\le N}\mu_j\mu_k\gamma_{jk}\right], \eqno(2.32a)$$
 $$f^\prime=\sum_{\mu=0,1}{\rm exp}\left[\sum_{j=1}^N\mu_j\left(\xi_j+d_j-{\pi\over 2}\,{\rm i}\right)
+\sum_{1\le j<k\le N}\mu_j\mu_k\gamma_{jk}\right], \eqno(2.32b)$$
  $$g=\sum_{\mu=0,1}{\rm exp}\left[\sum_{j=1}^N\mu_j\left(\xi_j-d_j+{\pi\over 2}\,{\rm i}\right)
+\sum_{1\le j<k\le N}\mu_j\mu_k\gamma_{jk}\right], \eqno(2.33a)$$
 $$g^\prime=\sum_{\mu=0,1}{\rm exp}\left[\sum_{j=1}^N\mu_j\left(\xi_j-d_j-{\pi\over 2}\,{\rm i}\right)
+\sum_{1\le j<k\le N}\mu_j\mu_k\gamma_{jk}\right], \eqno(2.33b)$$
where
$$\xi_j=p_jy+{1\over p_j}\tau+\xi_{j0}, \qquad (j=1, 2, ..., N),\eqno(2.34a)$$
$${\rm e}^{\gamma_{jk}}=\left({p_j-p_k\over p_j+p_k}\right)^2, \qquad (j, k=1, 2, ..., N; j\not=k),
\eqno(2.34b)$$
$$e^{d_j}=\sqrt{1+{\rm i}p_j\over 1-{\rm i}p_j},\qquad (j=1, 2, ..., N).\eqno(2.34c)$$
Here, $p_j$ and $\xi_{j0}$ are arbitrary complex  parameters satisfying the conditions $p_j\not=\pm p_k$
for $j\not= k$, ${\rm i}=\sqrt{-1}$ and $N$ is an arbitrary positive integer. The notation $\sum_{\mu=0,1}$
implies the summation over all possible combination of $\mu_1=0, 1, \mu_2=0, 1, ..., 
\mu_N=0, 1$}.\par 
\medskip
\noindent{\bf Proof.} It has been shown that $f$ and $f^\prime$ from (2.32) and $g$ and $g^\prime$
from (2.33)  satisfy the
bilinear equations (2.15) and (2.16), respectively [6]. 
To prove that (2.32) and (2.33) satisfy the bilinear equations (2.22) and (2.23), we replace the parameters $p_j$ by $-{\rm i}p_j\ (j=1, 2, ..., N)$ 
and shift the phase parameters
as $\xi_{j0} \rightarrow \xi_{j0}-d_j \ (j=1, 2, ..., N)$ and at the same time 
change the variables $y$ and $\tau$ as $y\rightarrow {\rm i}y, \tau\rightarrow -{\rm i}\tau$, respectively. 
It then turns out that $\xi_j$ and $\gamma_{jk}$ defind respectively by (2.34a) and (2.34b) remain the same form.
Thus, the proof reduces to the theorem 2.2 in I.
 See also  Remark 2.5 in I. \hspace{\fill}$\square$ \par
 \medskip
  The parametric solution (2.29) with (2.32) and (2.33) is characterized by the $2N$ complex parameters $p_j$ and $\xi_{j0}\ (j=1, 2, ..., N)$. It produces
 in general the complex-valued solutions.  The real-valued solutions are obtainable if one imposes certain conditions on these parameters. 
 Actually, there  arise various type of solutions depending on values of the parameters. These solutions include kinks, antikinks and breathers.
 Among them, we consider following three types: \par
 \medskip
 \noindent {\it Type 1: Kink solution} \par
 \noindent First, let $p_j$ and $\xi_{j0}\ (j=1, 2, ..., N)$ be real quantities. It then follows from (2.34c) that $d_j^*=-d_j\ (j=1, 2, ..., N)$ 
 where the asterisk denotes the complex conjugate.
 Taking account this fact in (2.32) and (2.33), we find that $f^\prime=g^*$ and   $g^\prime=f^*$. 
 These relations are substituted into (2.29) to give the real parametric  solution
 $$u(y,\tau)={\rm i}\,\ln{f^* g^* \over fg}, \eqno(2.35a)$$
  $$x(y,\tau)=y-\tau + {\rm i}\,\ln{f^* g\over fg^*} + y_0. \eqno(2.35b)$$
 As will be demonstrated in section 3, this solution yields the $N$-kink solution if the parameters  $p_j$  
 are all positive. The expression $u_x$ represents the $N$-soliton solution which decays at infinity. Remarkably, the amplitude of each soliton
 does not exceed 1 by virtue of the relation (2.31).
 The $N$-antikink solution is also available
  if these parameters are set to be negative. 
 We can also produce kink-antikink solutions. For example, let $N=M+M^\prime$ where $M$ and $M^\prime$ are positive
 integers and assign $M$ positive and $M^\prime$ negative values for $N$ parameters $p_j\ (j=1, 2, ..., N)$. Then, the parametric solution represents 
 the solution describing the interaction among $M$ kinks and $M^\prime$ antikinks.
 \par
 \medskip
 \noindent {\it Type 2: Breather solution} \par
 \noindent The second example is the parameterization which gives rise to the breather solution. To show this, we put
 $N=2M$ where  $M$ is a positive integer, and specify the parameters $p_j$ and $\xi_{j,0}\ (j=1, 2, ..., 2M)$ as
 $$p_{2j-1}=p_{2j}^*, \qquad \xi_{2j-1,0}=\xi_{2j,0}^*,\quad (j=1, 2, ..., M). \eqno(2.36)$$
 We see from (2.34) and (2.36) that $\xi_{2j-1}=\xi_{2j}^*,\ d_{2j-1}=-d_{2j}^*\ (j=1, 2, ..., M)$.
 It  turns out from (2.32) and (2.33) that $f^\prime=g^*$ and   $g^\prime=f^*$. Then, the  solution can be written in the same form as (2.35). \par
 \medskip
 \noindent {\it Type 3: Kink-breather solution} \par
 \noindent Let $N=2M+M^\prime$ where $M$ and $M^\prime$ are positive integers. In addition to the parameterization given by (2.36), 
 the $2M^\prime$ parameters $p_j(>0)$ and $\xi_{j0}\ (j=2M+1, 2M+2, ..., 2M+M^\prime)$ are chosen to be real. Then, the parameteric solution 
 (2.35) represents the solution describing the interaction among $M$ breathers and $M^\prime$ kinks.  
 The antikink-breather solution can be constructed similarly.
 \par
 For the above three types of solutions, $\phi$ from (2.18b) and $u_x$ from (2.31) can be given explicitly in terms of
 the tau functions $f, g$ and their complex conjugate as 
     $$\phi=\ln\,{g^*g\over f^*f}, \eqno(2.37)$$
   $$u_x={(g^*g)^2-(f^*f)^2\over (g^*g)^2+(f^*f)^2}. \eqno(2.38)$$
   Note that (2.37) provides real solutions of equation (2.11). \par
   \bigskip
   
   \leftline{\bf 3. Properties of solutions}\par
  \bigskip
  \noindent In this section, we describe the properties of real solutions constructed in section 2. We address both the kink and
  breather solutions. 
  \par
 \medskip
    \leftline{\it 3.1. 1-soliton solutions}\par
    \medskip
 \noindent  The tau-functions for the 1-soliton solutions are given by (2.32)and (2.33) with $N=1$:
  $$f=1+{\rm i}{\rm e}^{\xi_1+d_1},  \eqno(3.1a)$$
  $$g=1+{\rm i}{\rm e}^{\xi_1-d_1}, \eqno(3.1b)$$
  with
  $$\xi_1=p_1y+{\tau\over p_1}+\xi_{10}, \qquad{\rm e}^{ d_1}=\sqrt{1+{\rm i}p_1\over 1-{\rm i}p_1}. \eqno(3.1c)$$
  The real parameters $p_1$ and $\xi_{10}$ are related to the amplitude and phase of the soliton, respectively and $\xi_1$
  is the phase variable characterizing  the solution. 
 The parametric representation of the solution (2.35) can be written in the form
  $$u=2\,\tan^{-1}\left(\sqrt{1+p_1^2}\,\sinh\,\xi_1\right)+\pi,\eqno(3.2a)$$
  $$x=y-\tau+2\,\tan^{-1}(p_1\tanh\,\xi_1)+2\,\tan^{-1}p_1+y_0. \eqno(3.2b)$$
      Note that if $u$ solves equation (1.2), then so do the functions $\pm u+2\pi n\ (n : {\rm integer})$.
      For investigating solutions of traveling-wave type like 1-soliton solutions, it is convenient to parameterize solutions in terms of single
  variable $\xi_1$. To this end, we introduce a new variable $X$ by
  $$X\equiv x+c_1t+x_0={\xi_1\over p_1}+2\,\tan^{-1}(p_1\tanh\,\xi_1)+y_0, \eqno(3.3a)$$
     where
   $$c_1={1\over p_1^2}+1, \eqno(3.3b)$$
   and $x_0=\xi_{10}/p_1-2\,\tan^{-1}p_1$.
   Here, we used (2.3) and (3.2b).
    Observing the soliton in the original $(x,t)$ coordinate system,  it travels to 
   the left at the constant velocity $c_1$. 
   The soliton takes the form of a kink or an antikink depending on the sign of $p_1$. 
   To see this, we compute $u_X$ by using (3.2a) and (3.3a) to obtain
   $$u_X={2p_1\over \sqrt{1+p_1^2}}{\cosh\,\xi_1\over \cosh^2\xi_1+{p_1^2\over 1+p_1^2}}. \eqno(3.4)$$
   This expression implies that if $p_1>0$, then the solution $u$ becomes a monotonically increasing function 
   of $X$ and has the boundary values $u(-\infty)=0, u(+\infty)=2\pi$.
   If $p_1<0$, on the other hand, it represents an antkink solution.
   Figure 1 shows a typical profile of the kink solution as a function of $X$ together with the corresponding profile of $v\equiv u_X$.
   To study the  propagation characteristic of the soliton, we derive the dependence of the soliton velocity on the amplitude.
   To this end, let $A(>0)$ be the amplitude of $v$. It follows from (3.4) that
   $$A={2|p_1|\sqrt{1+p_1^2}\over 2p_1^2+1}={2\sqrt{c_1}\over c_1+1}, \eqno(3.5)$$
   where, in passing to the last line, we used the relation (3.3b).
   Notice from (3.5) and $c_1>1$ by (3.3b) that $0<A<1$.
   Solving (3.5) for $c_1$ gives
   $$c_1={1\over A^2}[-A^2+2+2\sqrt{1-A^2}]. \eqno(3.6)$$
   We see from (3.6)  that the velocity of the soliton is a monotonically decreasing function of the amplitude. 
   In other words, the small soliton
   travels faster than the large soliton. 
   Note, however, that if one transforms to the laboratory coordinate system $(Z, T)$ defined by the relations $Z=x+t, T=x-t$, then
   the velocity $\hat c_1$ of the soliton $u_Z$ turns out to be a monotonically increasing function of the amplitude $\hat A$.
   Indeed, expression corresponding  to (3.6) becomes $\hat c_1=1-2/(\hat A^2+1), \ \hat A>1$. This feature is the same as that of the 
   sG soliton solution expressed in terms of the laboratory coordinate. The situation is different for  soliton
   solutions of equation (1.1) with $\nu=-1$, as detailed in I. In this case, the velocity would become a monotonically
   decreasing function of the amplitude for certain range of the amplitude parameter.\par
      In view of the importance of the 1-soliton solution as an elementary solution, we provide an alternative derivation
   of the solution in appendix. The derivation is simpler compared with that presented in section 2 and is used
   frequently in reducing PDEs to tractable ordinary differential equations (ODEs). \par
   \bigskip
       \centerline{\bf Figure 1} \par
     \bigskip
   \leftline{\it 3.2. 2-soliton solutions}\par
   \medskip
   \noindent The tau-functions  for the 2-soliton solutions read from (2.32) and (2.33) with $N=2$ in the form
   $$ f=1+{\rm i}\left({\rm e}^{\xi_1+d_1}+{\rm e}^{\xi_2+d_2}\right)-\delta {\rm e}^{\xi_1+\xi_2+d_1+d_2}, \eqno(3.7a)$$
   $$ g=1+{\rm i}\left({\rm e}^{\xi_1-d_1}+{\rm e}^{\xi_2-d_2}\right)-\delta {\rm e}^{\xi_1+\xi_2-d_1-d_2}, \eqno(3.7b)$$
   with
   $$ \xi_j=p_jy+{\tau\over p_j}+\xi_{j0}, \qquad {\rm e}^{d_j}=\sqrt{1+{\rm i}p_j\over 1-{\rm i}p_j}\quad (j=1, 2),
   \qquad \delta={(p_1-p_2)^2\over (p_1+p_2)^2}. \eqno(3.7c)$$
     The parametric solution (2.35) with (3.7) represents three types of solutions, 
   depending on values of the parameters $p_j$ and $\xi_{0j}\ (j=1, 2)$,
   i.e.,  kink-kink,  kink-antikink  and  breather solutions.  \par
      \bigskip
   \leftline{\it 3.2.1. Kink-kink solution}\par
   \medskip
   \noindent   If we specify $p_1$ and $p_2$ be positive and $\xi_{01}$ and $\xi_{02}$ be real, then the kink-kink solution is obtained.
       The solution represents the so-called $4\pi$ kink.
       In  figure 2a-c, we depict a typical profile of  $v(\equiv u_x)$ instead of $u$ for three different times. 
       It represents the interaction of two solitons with the amplitudes $A_1=0.38$ and $A_2=0.75$. 
    As evidenced from figure 2, a smaller soliton overtakes,
    interacts and emerges ahead of a larger soliton. This reflects the fact that the velocity of each soliton is a monotonically
    decreasing function of its amplitude (see (3.6)). 
       The general formula for the phase shift arising from the interaction of $N$ solitons will be given by (3.18) below.
    In particular, for $N=2$, it reads
    $$\Delta_1=-{1\over p_1}\ln\left({p_1-p_2\over p_1+p_2}\right)^2+4\,\tan^{-1}p_2, \eqno(3.8a)$$
     $$\Delta_2={1\over p_2}\ln\left({p_1-p_2\over p_1+p_2}\right)^2-4\,\tan^{-1}p_1. \eqno(3.8b)$$
     It can be verified from (3.8) that $\Delta_1>0$ and $\Delta_2<0$ for $0<p_1<p_2$.
     In the present example, formula (3.8) yields $\Delta_1=10.3$ and $\Delta_2=-4.2.$
     If one observes the interaction process in the laboratory coordinate system introduced in section 3.1, then
     one can see that 
      the larger (smaller) soliton suffers a positive (negative) phase shift after the interaction, 
            which is in accordance with the property of the sG 2-soliton solution
      written in terms of the laboratory coordinate.                        \par
     \bigskip 
      \centerline{\bf Figure 2 a-c} \par
         \bigskip
     \leftline{\it 3.2.2. Breather solution}\par
    \medskip
       \noindent The breather solution can be constructed following the parameterization given by (2.36). For $M=1$, let
         $$p_1=a+{\rm i}b, \qquad p_2=a-{\rm i}b=p_1^*,\qquad (a>0,\ b>0), \eqno(3.9a)$$
   $$\xi_{10}=\lambda+{\rm i}\mu, \qquad \xi_{20}=\lambda-{\rm i}\mu=\xi_{10}^*. \eqno(3.9b)$$
   Then, $f$ and $g$ from (2.32) and (2.33) become
   $$ f=1+{\rm i}({\rm e}^{\xi_1+d_1}+{\rm e}^{\xi_1^*-d_1^*})+\left({b\over a}\right)^2{\rm e}^{\xi_1+\xi_1^*+d_1-d_1^*}, \eqno(3.10a)$$
   $$ g=1+{\rm i}({\rm e}^{\xi_1-d_1}+{\rm e}^{\xi_1^*+d_1^*})+\left({b\over a}\right)^2{\rm e}^{\xi_1+\xi_1^*-d_1+d_1^*}, \eqno(3.10b)$$
      where 
      $$\xi_1=\theta+{\rm i}\chi, \eqno(3.10c)$$
$$\theta=a\left(y+{1\over a^2+b^2}\tau\right)+\lambda, \eqno(3.10d)$$
$$\chi=b\left(y-{1\over a^2+b^2}\tau\right)+\mu, \eqno(3.10e)$$
$${\rm e}^{d_1}=\sqrt{1-a^2-b^2+2{\rm i}a\over a^2+(1-b)^2}\equiv\alpha\,{\rm e}^{{\rm i}\beta}. \eqno(3.10f)$$
The tau functions  $f$ and $g$ can be written in terms of the  new variables defined by (3.10) as
$$f=1+\left\{-\alpha\,\sin(\chi+\beta)+{1\over \alpha}\,\sin(\chi-\beta)\right\}{\rm e}^\theta+\left({b\over a}\right)^2{\rm e}^{2\theta}\cos\,2\beta$$
$$+{\rm i}\left[\left\{\alpha\,\cos(\chi+\beta)+{1\over \alpha}\,\cos(\chi-\beta)\right\}{\rm e}^\theta+\left({b\over a}\right)^2{\rm e}^{2\theta}\sin\,2\beta\right], \eqno(3.11a)$$
$$g=1+\left\{\alpha\,\sin(\chi+\beta)-{1\over \alpha}\,\sin(\chi-\beta)\right\}{\rm e}^\theta+\left({b\over a}\right)^2{\rm e}^{2\theta}\cos\,2\beta$$
$$+{\rm i}\left[\left\{\alpha\,\cos(\chi+\beta)+{1\over \alpha}\,\cos(\chi-\beta)\right\}{\rm e}^\theta-\left({b\over a}\right)^2{\rm e}^{2\theta}\sin\,2\beta\right]. \eqno(3.11b)$$
The phase variable $\theta$ characterizes the envelope of the breather whereas the phase variable $\chi$ governs the internal oscillation.
The parametric solution can be written by (2.35).
           In figure 3a-c, a typical profile of $u$ is depicted for three different times. We see that the breather propagates to
   the left while changing its profile. The propagation characteristic of the breather is similar to that presented in I. \par
      \medskip
   \centerline{\bf Figure 3 a-c} \par
      \bigskip
        \leftline{\it 3.3. N-soliton solutions} \par
        \medskip
   \noindent The solutions including an arbitrary number of solitons can be constracted from the parametric representation (2.35) with tau functions (2.32) and (2.33). 
   There exist 
   a variety of solutions which are composed of any combination of kink, antikink and breather solutions. 
  Here, we address the $N$-kink solutions and $M$ breather solutions. 
    For the former solutions, we investigate the asymptotic behavior of solutions for
   large time and derive the formulas for the phase shift while for the latter ones, we provide a recipe for
   constructing $M$ breather solution from the $N$-soliton solution. As an example, we present a 
   solution describing the interaction between a kink and a breather. \par 
      \par
   \bigskip
   \leftline{\it 3.3.1.  N-kink solution} \par
   \medskip
   \noindent Let the velocity of the $j$th kink be $c_j=(1/p_j^2)+1\ (p_j>0)$
   and order the magnitude of the velocity of each kink
as $c_1>c_2> ...>c_N$.  We observe the interaction of $N$ kinks in a moving frame
with a constant velocity $c_n$. We  take the limit $t \rightarrow -\infty$
with the phase variable $\xi_n$ being fixed.  We then find that $f$ and $g$
have the following leading-order asymptotics
$$ f \sim \delta_{n}\,\exp\left[\sum_{j=n+1}^N\left(\xi_j+d_j+{\pi\over 2}{\rm i}\right)\right]
\left(1+{\rm i}{\rm e}^{\xi_n+d_n+\delta_n^{(-)}}\right), \eqno(3.12a)$$
$$ g \sim \delta_{n}\,\exp\left[\sum_{j=n+1}^N\left(\xi_j-d_j+{\pi\over 2}{\rm i}\right)\right]
\left(1+{\rm i}{\rm e}^{\xi_n-d_n+\delta_n^{(-)}}\right), \eqno(3.12b)$$
where
$$\delta_n^{(-)}=\sum_{j=n+1}^N\ln\left({p_n-p_j\over p_n+p_j}\right)^2, \eqno(3.12c)$$
$$\delta_{n}=\prod_{n+1\leq j<k\leq N}\left({p_j-p_k\over p_j+p_k}\right)^2. \eqno(3.12d)$$
If we substitute (3.12) into (2.35), we obtain the asymptotic form of $u$ and $x$:
 $$u\sim 2\,\tan^{-1}\left[\sqrt{1+p_n^2}\,\sinh\left(\xi_n+\delta_n^{(-)}\right)\right]+\pi,\eqno(3.13a)$$
  $$x\sim y-\tau+2\,\tan^{-1}\left[p_n\tanh\,\left(\xi_n+\delta_n^{(-)}\right)\right]+4 \sum_{j=n+1}^N\tan^{-1}p_j+2\,\tan^{-1}p_n+y_0. \eqno(3.13b)$$
As $t \rightarrow +\infty$, the expressions corresponding to (3.13) are given by
$$u\sim \,2\tan^{-1}\left[\sqrt{1+p_n^2}\,\sinh\left(\xi_n+\delta_n^{(+)}\right)\right]+\pi,\eqno(3.14a)$$
  $$x\sim y-\tau+2\,\tan^{-1}\left[p_n\tanh\,\left(\xi_n+\delta_n^{(+)}\right)\right]+4 \sum_{j=1}^{n-1}\tan^{-1}p_j+2\,\tan^{-1}p_n+y_0. \eqno(3.14b)$$
with
$$\delta_n^{(+)}=\sum_{j=1}^{n-1}\ln\left({p_n-p_j\over p_n+p_j}\right)^2. \eqno(3.14c)$$
\par
Let $x_c$ be the center position of the $n$th kink in the $(x,t)$ coordinate
system. It simply stems from the relation $\xi_n+\delta_n^{(\pm)}=0$ by invoking (3.13a)
and (3.14a). Thus, as  $t \rightarrow -\infty$
$$x_c+c_nt+x_{n0} \sim -{1\over p_n}\delta_n^{(-)}+4 \sum_{j=n+1}^N\tan^{-1}p_j+y_0, \eqno(3.15)$$
where $x_{n0}=\xi_{n0}/p_n-2\,\tan^{-1}p_n$.
 As $t \rightarrow +\infty$, on the other hand,
the corresponding expression turns out to be
$$x_c+c_nt+x_{n0} \sim -{1\over p_n}\delta_n^{(+)}+4 \sum_{j=1}^{n-1}\tan^{-1}p_j+y_0. \eqno(3.16)$$
If we take into account the fact that all kinks propagate to the left, we can
define the phase shift of the $n$th kink as
$$\Delta_n=x_c(t\rightarrow -\infty)-x_c(t\rightarrow +\infty). \eqno(3.17)$$
Using (3.12c), (3.14c), (3.15) and (3.16), we find that
$$\Delta_n={1\over p_n}\left\{\sum_{j=1}^{n-1}\ln\left({p_n-p_j\over p_n+p_j}\right)^2
-\sum_{j=n+1}^N\ln\left({p_n-p_j\over p_n+p_j}\right)^2\right\}$$
$$+4 \sum_{j=n+1}^N\tan^{-1}p_j- 4 \sum_{j=1}^{n-1}\tan^{-1}p_j,
\quad (n = 1, 2, ..., N). \eqno(3.18)$$
The first term on the right-hand side of (3.18)  coincides with the formula
for the phase shift arising from the interaction of $N$ kinks of the
sG equation [6, 7, 9] whereas the second and third terms appear as a consequence of the coordinate
transformation (2.3).  \par
    \bigskip
   \leftline{\it 3.3.2.  M-breather solution} \par
   \medskip
   \noindent  The construction of the $M$-breather solution can be done following the similar procedure  to that for the
   1-breather solution developed in section 3.2.2.
    To proceed, we specify the parameters in (2.32) and (2.33) for the tau-functions $f$ and $g$ as
$$p_{2j-1}=p_{2j}^*\equiv a_j+{\rm i}b_j,\quad a_j>0,\quad b_j>0,
\quad (j=1, 2, ..., M),\eqno(3.19a)$$
$$\xi_{2j-1,0}=\xi_{2j,0}^*\equiv \lambda_j+{\rm i}\mu_j,\quad (j=1, 2, ..., M).\eqno(3.19b)$$
Then, the phase variables $\xi_{2j-1}$ and $\xi_{2j}$ are written as
$$\xi_{2j-1}=\theta_j+{\rm i}\chi_j, \quad (j=1, 2, ..., M),\eqno(3.20a)$$
$$\xi_{2j}=\theta_j-{\rm i}\chi_j, \quad (j=1, 2, ..., M),\eqno(3.20b)$$
with the real phase variables
$$\theta_j=a_j(y+c_j\tau)+\lambda_j, \quad (j=1, 2, ..., M), \eqno(3.20c)$$
$$\chi_j=b_j(y-c_j\tau)+\mu_j, \quad (j=1, 2, ..., M), \eqno(3.20d)$$
$$c_j={1\over a_j^2+b_j^2}, \quad (j=1, 2, ..., M). \eqno(3.20e)$$
\par
The parametric solution (2.35) with  (3.19) and (3.20) describes
multiple collisions of $M$  breathers.
One can perform an asymptotic analysis for the $M$-breather solution, showing that the $M$-breather solution splits into $M$ single
   breathers as $t\rightarrow \pm\infty$.
     The resulting asymptotic form of the solution is, however,  too complicated to write down and hence we omit the detail.
   One can refer to the similar analysis to that for the $M$-breather solution of the short pulse equation [10].\par
   \bigskip
     \leftline{\it 3.3.3 Kink-breather solution}\par
   \medskip
   \noindent  We take a 3-soliton solution with parameters  $p_j$ and $\xi_{0j}$ $(j=1, 2, 3)$.
   If one impose the conditions that $p_2=p_1^*, \xi_{02}=\xi_{01}^*$ as  already specified for the breather solution (see section 3.2.2) 
   and $p_3(>0),\ \xi_{03}$ real for the  kink solution, then
   the expression of $u$ would represent a solution describing the interaction between a kink and a breather.
   We choose $p_1, p_2, \xi_{10}$ and $\xi_{20}$ as those given by (3.9). Then, the tau functions $f$ and $g$ from (2.32) and (2.33)
   become
   $$f=1+{\rm i}\left(s_1{\rm e}^{\xi_1}+{1\over s_1^*}{\rm e}^{\xi_1^*}+s_3{\rm e}^{\xi_3}\right)+\left({b\over a}\right)^2{s_1\over s_1^*}{\rm e}^{\xi_1+\xi_1^*}$$
   $$-\delta_{13}s_1s_3{\rm e}^{\xi_1+\xi_3}-\delta_{13}^*{s_3\over s_1^*}{\rm e}^{\xi_1^*
   +\xi_3}+{\rm i}\left({b\over a}\right)^2{s_1s_3\over s_1^*}\delta_{13}\delta_{13}^*{\rm e}^{\xi_1+\xi_1^*+\xi_3}, \eqno(3.21a)$$
   $$g=1+{\rm i}\left({1\over s_1}{\rm e}^{\xi_1}+ s_1^*{\rm e}^{\xi_1^*}+{1\over s_3}{\rm e}^{\xi_3}\right)+\left({b\over a}\right)^2{s_1^*\over s_1}{\rm e}^{\xi_1+\xi_1^*}$$
   $$-{\delta_{13}\over s_1s_3}{\rm e}^{\xi_1+\xi_3}-\delta_{13}^*{s_1^*\over s_3}{\rm e}^{\xi_1^*
   +\xi_3}+{\rm i}\left({b\over a}\right)^2{s_1^*\over s_1s_3}\delta_{13}\delta_{13}^*{\rm e}^{\xi_1+\xi_1^*+\xi_3}. \eqno(3.21b)$$
     where
   $$s_1={\rm e}^{d_1}=\sqrt{1-b+{\rm i}a\over 1-b-{\rm i}a}={1\over s_2^*},\qquad s_3=\sqrt{1+{\rm i}p_3\over 1-{\rm i}p_3},
   \qquad \delta_{13}=\left({a-p_3+{\rm i}b\over a+p_3+{\rm i}b}\right)^2=\delta_{23}^*. \eqno(3.21c)$$
   \par
   Figure 4a-c shows a typical profile of $v\equiv u_x$ for three different times. 
  We see that the soliton overtakes the breather whereby it suffers a phase shift.
  An asymptotic analysis using the tau functions (3.21) yields the formula for the phase
  shift of the soliton, which we denote $\Delta$. Actually,  one has for $p_3^2<a^2+b^2$
  $$\Delta={2\over p_3}\, \ln{(p_3+a)^2+b^2\over (p_3-a)^2+b^2}+ 4\,\tan^{-1}{2a\over 1-a^2-b^2}, \eqno(3.22a)$$
  and for $a^2+b^2<p_3^2$
  $$\Delta=-{2\over p_3}\, \ln{(p_3+a)^2+b^2\over (p_3-a)^2+b^2}- 4\,\tan^{-1}{2a\over 1-a^2-b^2}. \eqno(3.22b)$$
  In the present example, formula (3.22a) gives $\Delta=7.7$. \par
       \bigskip
   \centerline{\bf Figure 4 a-c}\par
   \bigskip
    \leftline{\it 3.3.4 Breather-breather solution}\par
    \noindent The breather-breather (or 2-breather) solution is reduced from a 4-soliton solution following the procedure
   described in section 3.3.2. Figure 5a-c shows  a typical profile of $u$ for three different times. 
   It represents a typical feature common to the interaction of solitons, i.e., each breather recovers its profile
   after collision.  
   \par
    \medskip
   \centerline{\bf Figure 5a-c}\par
   \bigskip
              \par\leftline{\bf 4. Reduction to the short pulse and sG equations} \par
   \bigskip
   \noindent We write the short pulse equation in the form
   $$u_{tx}=u-{\nu\over 6 }(u^3)_{xx}, \eqno(4.1)$$
   where $u=u(x,t)$ represents the magnitude of the electric field and $\nu$ is a real
   constant.
       The short pulse   equation (4.1)  with $\nu=-1$ was proposed as a model nonlinear equation 
describing the propagation of ultra-short optical pulses in nonlinear media [11]. 
Quite recently, equation (4.1) with $\nu=1$
 $$u_{tx}=u-{1\over 6 }(u^3)_{xx}, \eqno(4.2)$$
 was shown to model the evolution of ultra-short pulses in the band gap of nonlinear metamaterials [12].
 Here, we demonstrate that the generalized
sG equation (1.2)  is reduced to an alternative version of the  short pulse equation (4.2)  by taking an appropriate scaling limit combined with a
coordinate transformation. The  reduced forms of equations corresponding to (2.7), (2.9), (2.10) and (2.11) are also presented.
The $N$-soliton solution of the short pulse equation can be derived from
that of the generalized sG equation.  The reduction to the sG equation is briefly discussed. \par
\bigskip
\leftline{\it 4.1. Reduction to the short pulse equation} \par
\leftline{\it 4.1.1. Scaling limit of the generalized sG equation}\par
\medskip
 \noindent The reduction to the short pulse equation (4.2) can be done by employing the procedure developed in I. Therefore, we outline the result.
 Let us first introduce new variables with bar according to
the relations
$$\bar u={u\over \epsilon},\qquad \bar x={1\over\epsilon}(x+t),\qquad \bar y={y\over \epsilon} \qquad  \bar y_0={y_0\over\epsilon},
\qquad \bar t=\epsilon t, \qquad \bar\tau=\epsilon \tau,$$
   $$\qquad \bar p_j=\epsilon p_j, \qquad \bar\xi_{j0}= \xi_{j0},\qquad (j=1, 2, ..., N), \eqno(4.3)$$ 
   where $\epsilon$ is a small parameter and the quantities with bar are assumed to be order 1. Rewriting equation (1.2)
    in terms of the new variables and expanding $\sin\,\epsilon \bar u$ in an
   infinite series with respect to $\epsilon$ and  comparing terms of order $\epsilon$ on both sides, we obtain  
   equation (4.2) written by  the new variables. \par 
   Under the scaling (4.3), expression (2.7) is invariant and hence we put $\bar \phi=\phi$ to give
   $$\bar u_{\bar y}=\sinh\,\bar\phi. \eqno(4.4)$$
   Equation (2.9) then reduces to
   $$\bar\phi_{\bar\tau}=\bar u. \eqno(4.5)$$
   Equations (2.10) and (2.11) now become
      $${\bar u_{\bar\tau \bar y}\over \sqrt{1+{\bar u_{\bar y}}^2}}=
      \bar u \eqno(4.6)$$
  $$\bar\phi_{\bar \tau \bar y}=\sinh\,\bar\phi, \eqno(4.7)$$
   respectively. Equation (4.7) is known as the sinh-Gordon equation. \par
         \bigskip
   \leftline{\it 4.1.2. Scaling limit of the N-soliton solution}\par
   \medskip
   \noindent  To derive the scaling limit of the $N$-soliton solution, we use the expansion 
   $${\rm exp}\left(\sum_{j=1}^N\mu_jd_j\right)=\prod_{j=1}^N\left({1+{{\rm i}p_j\over {\epsilon}}\over 1-{{\rm i}p_j\over {\epsilon}}}\right)^{\mu_j\over 2}
   = {\rm exp}\left({\pi\over 2} {\rm i}\sum_{j=1}^N\mu_j\right)\left(1-\epsilon \sum_{j=1}^N{\mu_j\over \bar p_j}\right)+O(\epsilon^2). \eqno(4.8)$$
   as well as the scaled variables (4.3). These are substituted into (2.32a) to obtain the expansion of the tau function $f$
   \begin{alignat*}{1}
    f &= \sum_{\mu=0, 1}\left(1-{\rm i}\epsilon \sum_{j=1}^N{\mu_j\over \bar p_j}\right)
   {\rm exp}\left[\sum_{j=1}^N\mu_j\left(\bar\xi_j+\pi{\rm i}\right)
+\sum_{1\le j<k\le N}\mu_j\mu_k\bar\gamma_{jk}\right]+O(\epsilon^2) \\
&=\bar f-{\rm i}\epsilon\bar f_{\bar\tau}+O(\epsilon^2), \tag{4.9a}
 \end{alignat*}
where 
$$\bar f=\sum_{\mu=0,1}{\rm exp}\left[\sum_{j=1}^N\mu_j\left(\bar\xi_j+\pi{\rm i}\right)
+\sum_{1\le j<k\le N}\mu_j\mu_k\bar\gamma_{jk}\right], \eqno(4.9b)$$
 $$\bar\xi_j=\bar p_j\bar y+{\bar\tau\over\bar p_j}+\bar\xi_{j0}, \qquad (j=1, 2, ..., N),\eqno(4.9c)$$
   $${\rm e}^{\bar\gamma_{jk}}=\left({\bar p_j-\bar p_k\over \bar p_j+\bar p_k}\right)^2, \qquad (j, k=1, 2, ..., N; j\not=k).
\eqno(4.9d)$$
   Similarly, it follows from (2.32b), (2.32c), (2.32d) and (4.8) that
   $$f^\prime=\bar g-{\rm i}\epsilon\bar g_{\bar\tau}+O(\epsilon^2). \eqno(4.10a)$$
   $$g =\bar g+{\rm i}\epsilon\bar g_{\bar\tau}+O(\epsilon^2). \eqno(4.10b)$$
   $$g^\prime=\bar f+{\rm i}\epsilon\bar f_{\bar\tau}+O(\epsilon^2), \eqno(4.10c)$$
   with
   $$\bar g=\sum_{\mu=0,1}{\rm exp}\left[\sum_{j=1}^N\mu_j\bar\xi_j
+\sum_{1\le j<k\le N}\mu_j\mu_k\bar\gamma_{jk}\right]. \eqno(4.10d)$$
   By introducing (4.3), (4.9) and (4.10) into (2.29) and taking the limit $\epsilon\rightarrow 0$, we obtain the parametric solution
   of the short pulse equation (4.2) in terms of the tau functions $\bar f$ and $\bar g$ as follows:
   $$\bar u=2\left(\ln\,{\bar g\over \bar f}\right)_{\bar \tau}, \eqno(4.11a)$$
   $$\bar x=\bar y-2(\ln\,\bar f \bar g)_{\bar\tau}+\bar y_0. \eqno(4.11b)$$
   If the tau functions are real, then (4.11) gives rise to real solutions. The construction of real solutions can be
   done by means of the procedure developed for the short pulse equation [10, 13, 14]. \par
   We exemplify real solutions by taking all parameters be real.
  It turns out that the solution develops singularities.  We can confirm this fact  by considering the simplest
   1-soliton solution. In fact, it follows from (4.9)  and (4.10) that the corresponding tau functions are given by
   $$f=1-{\rm e}^{\xi_1}, \qquad g=1+{\rm e}^{\xi_1}, \qquad \xi_1=p_1y+{\tau\over p_1}+\xi_{10}, \eqno(4.12)$$
   where the bar appended to the variables is omitted for simplicity. The parametric solution (4.11) now  becomes
   $$u=-{2\over p_1}{1\over \sinh\,\xi_1}, \eqno(4.13a)$$
   $$x=y-{2\over p_1}{1\over \tanh\,\xi_1}-{2\over p_1}+y_0. \eqno(4.13b)$$
   Although $u_x$ takes finite
    value as expected from the scaling limit of the relation (2.31), $u$ itself diverges as
   $|x|\rightarrow \infty$. The general $N$-soliton solution exhibits the singular nature. 
    Note from (4.5) and (4.11a) that $\bar\phi=2\,{\rm ln}(\bar g/\bar f)$ gives the singular $N$-soliton
   solution of the sinh-Gordon equation (4.7) [15]. The more detailed description of solutions will be
   reported elsewhere.
     \par
  \bigskip
\leftline{\it 4.2. Reduction to the sG equation}\par
\medskip
\noindent If we introduce the following new scaled variables
$$\bar u=u,\qquad \bar x=\epsilon x,\qquad \bar y=\epsilon y,\qquad \bar t={t\over \epsilon},\qquad \bar \tau={\tau\over \epsilon},$$
$$\bar p_j={p_j\over \epsilon},\qquad \bar \xi_{j0}=\xi_{j0},\ (j=1, 2, ..., N), \eqno(4.14)$$
then in the limit of $\epsilon \rightarrow 0$, we can deduce the generalized sG equation (1.2) to the sG equation 
$$\bar u_{\bar t\bar x}=\sin \bar u. \eqno(4.15)$$
The scaling limit of (2.29b) now 
 leads to the expression $\bar y=\bar x$ which, combined with the obvious relation
$\bar\tau=\bar t$, yields the limiting form of the tau functions (2.32) and (2.33)
$$f=\bar f, \qquad f^\prime=\bar f^\prime, \qquad  g=\bar f, \qquad g^\prime=\bar f^\prime, \eqno(4.16a)$$
where
 $$\bar f=\sum_{\mu=0,1}{\rm exp}\left[\sum_{j=1}^N\mu_j\left(\bar\xi_j+{\pi\over 2}\,{\rm i}\right)
+\sum_{1\le j<k\le N}\mu_j\mu_k\bar\gamma_{jk}\right], \eqno(4.16b)$$
 $$\bar f^\prime=\sum_{\mu=0,1}{\rm exp}\left[\sum_{j=1}^N\mu_j\left(\bar\xi_j-{\pi\over 2}\,{\rm i}\right)
+\sum_{1\le j<k\le N}\mu_j\mu_k\bar\gamma_{jk}\right], \eqno(4.16c)$$
$$\bar\xi_j=\bar p_j\bar x+{\bar t\over\bar p_j}+\bar\xi_{j0}, \qquad (j=1, 2, ..., N), \eqno(4.16d)$$
$${\rm e}^{\bar\gamma_{jk}}=\left({\bar p_j-\bar p_k\over \bar p_j+\bar p_k}\right)^2, \qquad (j, k=1, 2, ..., N; j\not=k).
\eqno(4.16e)$$
 The parametric solution (2.29) with the tau functions (2.32) and (2.33) reduces to the usual form of the $N$-soliton solution of the sG equation i.e., 
$$\bar u(\bar x,\bar t)=2{\rm i}\, \ln\,{\bar f^\prime\over \bar f}. \eqno(4.17)$$
\par
As for the scaling limit of the amplitude, the appropriate scaling is $A=\epsilon\bar A$. This relation and the scaling $p_1=\epsilon\bar p_1$ from (4.14)
are introduced into (3.5) to obtain the limiting relation $\bar A=2\bar p_1,\ \bar p_1>0$, which is just the amplitude of the sG
soliton in the $(x, t)$ coordinate system. 
The phase shift is scaled by $\bar\Delta_n=\epsilon\Delta_n$. The limiting form of the phase shift is given by the
first term on the right-hand side of (3.18), reproducing the formula for the $N$-soliton solution of the sG equation [6, 8, 9]. 
Thus, the sG limit can be performed consistently.
\par
 \bigskip
\leftline{\bf 5. Conservation laws}\par
\noindent The generalized sG equation (1.2) possesses an infinite number of conservation laws. Their construction
can be done following the similar procedure to that developed for equation (1.1) with $\nu=-1$ [2].
Here, we shall demonstrate it shortly.\par
 First, let
$$\sigma=u-{\rm i}\, \sinh^{-1}u_y. \eqno(5.1)$$
By  direct substitution, we find the relation
$$\sigma_{\tau y}-\sin\, \sigma=\left\{(1+u_y^2)^{1\over 2}-{\rm i}\,{\partial\over\partial y}\right\}
\left\{{u_{\tau y}\over (1+u_y^2)^{1\over 2}}-\sin\,u\right\}. \eqno(5.2)$$
Thus, if $u$ is a solution of equation (2.10), then $\sigma$ given by (5.1) satisfies the sG equation (2.13a).
Recall that equation (2.10) is a transformed form of an integrable  equation (1.2) by means of the hodograph transformation (2.3). Thus, relation (5.1) gives
a B\"acklund transformation between solutions $u$ and $\sigma$ of the two integrable equations. This observation allows us 
to obtain conservation laws of equation (1.2) quite simply. First,
  note that the sG equation (2.13a) admits local conservation laws of the form [16, 17]
$$P_{n,\tau}=Q_{n,y},\qquad (n=0, 1, 2, ...), \eqno(5.3)$$
where $P_n$ and $Q_n$ are polynomials of $\sigma$ and its $y$-derivatives. Rewriting this relation in terms of the original variables $x$ and $t$ by (2.4) and
using equation (2.2), we can recast (5.3) to the form
$$(rP_n)_t=(rP_n\cos\,u+Q_n)_x. \eqno(5.4)$$
The quantities 
$$I_n=\int^\infty_{-\infty}rP_ndx,\qquad (n=0, 1, 2, ...), \eqno(5.5)$$
 then become the conservation laws of equation (1.2)
upon  substitution of (5.1). The explicit calculation of conservation laws can be done straightforwardly. We present
the first three of them. The corresponding $P_n$ for the sG equation may be written as [16, 17]
$$P_0=1-\cos\,\sigma,\qquad P_1={1\over 2}\sigma_y^2, \qquad P_2={1\over 4}\sigma_y^4-\sigma_{yy}^2. \eqno(5.6)$$
It follows from (5.5), (5.6) and the relations $r_x=-u_xu_{xx}/r, (u_x/r)_x=u_{xx}/r^3$ which stem from (2.1) that 
$$I_0=\int^\infty_{-\infty}(r-\cos\,u)dx,\eqno(5.7a)$$
$$I_1={1\over 2}\int^\infty_{-\infty}\left({u_x^2\over r}-{u_{xx}^2\over r^5}\right)dx, \eqno(5.7b)$$
$$I_2=\int^\infty_{-\infty}\left[{1\over 4}{u_x^4\over r^3}+{3\over 2}{u_{xx}^2\over r^5}+{1\over r^7}\left(u_{xxx}^2-{5\over 2}u_{xx}^2\right)
+{7u_{xx}^4\over r^9}-{35\over 4}{u_{xx}^4\over r^{11}}\right]dx. \eqno(5.7c)$$
Note that terms including the imaginary unit i do not appear in (5.7) which take the form of total differential and
 become zero after integration with respect to $x$. \par
The conservation laws generated by the  procedure outlined above reduce to 
those of the short pulse and sG equations in the scaling limits described in section 4.
In particular, the first three conservation laws of the short pulse equation (4.2) read
$$I_0=\int^\infty_{-\infty}(r-1)dx,\eqno(5.8a)$$
$$I_1=-{1\over 2}\int^\infty_{-\infty}{u_{xx}^2\over r^5}dx, \eqno(5.8b)$$
$$I_2=\int^\infty_{-\infty}\left({u_{xxx}^2\over r^7}
+{7u_{xx}^4\over r^9}-{35\over 4}{u_{xx}^4\over r^{11}}\right)dx. \eqno(5.8c)$$
\par
\bigskip
\leftline{\bf 6. Conclusion}\par
\noindent In this paper, we have developed a systematic procedure for solving the
generalized sG equation (1.2). The structure of solutions was found to differ substantially 
from that of the generalized sG equation (1.1) with $\nu=-1$ which has been detailed in I.
We have presented three types of solutions, i.e., kink, breather and kink-breather solutions
and investigated their properties. We emphasize that equation (1.2) does not admit
singular solutions in the sense that solutions must have finite slope as required by the relation (2.31).
Consequently, loop solitons and other types of multi-valued solutions never exist. \par
The existence of multisoliton solutions and an infinite number of conservation laws strongly
support the complete integrability of the equation although its rigorous proof must be discussed
in a different mathematical context.  Another interesting issue to be resolved in a future work will be
the initial value problem. In the case of equation (1.1) with $\nu=-1$, the solution to the
problem can be expressed by the solution of a matrix Riemann-Hilbert problem [2].
At present, however, whether the method employed in [2] works well or not
for equation (1.2) is not known. As for solutions,
the bilinear formalism used here and in I can be applied to equation (1.1) as well to
obtain periodic solutions. Actually, some periodic solutions have been presented for the short pulse equation (4.19) with $\nu=-1$ [13, 14].
We expect that the generalized sG equation exhibits a variety of periodic solutions when compared with those of the sG equation.
These problems are currently under study. \par

\newpage
\leftline{\bf Appendix A. An alternative derivation of the 1-soliton solutions} \par
\bigskip
\noindent The 1-soliton solutions take the form of traveling wave
$$u=u(X), \qquad X=x+c_1t+x_0. \eqno(A.1)$$
Substituting this expression into equation (1.2) and integrating the resultant ODE  once with
respect to $X$  under the boundary condition $u(-\infty)=0\ ({\rm mod}\, 2\pi)$, we obtain
$$u_X^2={(c_1-\cos u)^2-(c_1-1)^2\over (c_1-\cos u)^2}. \eqno(A.2)$$
Since $u_X^2\geq 0$, we must require that the right-hand side of (A.2) is nonnegative.  
 One can see that this condition becomes $c_1\geq \cos^2(u/2)$.
In accordance with (3.3b), we solve equation (A.2) under the condition $c_1>1$. 
To proceed, we define a new variable $\xi$ by
$$X=\int(c_1-\cos u )d\xi. \eqno(A.3)$$
Then,  equation (A.2) reduces to
$$u_\xi=\pm\sqrt{(c_1-\cos u)^2-(c_1-1)^2}. \eqno(A.4)$$
Equation (A.4) is integrated through the change of the variable   $s=\tan(u/2)$.
After a few calculations, we obtain
$$s=\pm\sqrt{c_1-1\over c_1}{1\over \sinh\sqrt{c_1-1}\xi} \eqno(A.5)$$
and
$$\cos u={1-s^2\over 1+s^2}=1-{2(c_1-1)\over c_1\, \sinh^2\sqrt{c_1-1}\,\xi+c_1-1}. \eqno(A.6)$$
Substituting (A.6) into (A.3) and performing the integration with respect to $\xi$, we find
$$X=(c_1-1)\xi+2\,\tan^{-1}\left({1\over \sqrt{c_1-1}}\tanh(\sqrt{c_1-1}\,\xi\right)+y_0, \eqno(A.7)$$
where $y_0$ is an integration constant. It follows from (A.5) and the boundary condition for $u$ that
$$u=2\,\tan^{-1}\left(\sqrt{c_1\over c_1-1}\sinh\sqrt{c_1-1}\,\xi\right) +\pi. \eqno(A.8)$$
If we put
$c_1=(1/ p_1^2)+1\ (p_1>0)$ and $ \xi=p_1\xi_1$, then
we can see that (A.7) and (A.8) coincide with (3.3) and (3.2a), respectively. \par

\newpage
\leftline{\bf References}\par
\begin{enumerate}[{[1]}]
\item Fokas AS 1995 { On a class of physically important integrable equations} {\it Phys. D} {\bf 87} 145
\item Lenells J and Fokas AS 2009 { On a novel integrable generalization of the sine-Gordon equation} arXiv: 0909.2590v1[nlin. SI]
\item Matsuno Y 2010 { A direct method for solving the generalized sine-Gordon equation} {J. Phys. A: Math. Theor.} {\bf 43} 105204(28pp)
\item Hirota R 1980 { Direct Methods in Soliton Theory}
{\it Solitons} ed RK Bullough and DJ Caudrey 
({\it Topics in Current Physics 
{\rm vol. 17}}) (New York: Springer) p 157
\item Matsuno Y 1984 {\it Bilinear Transformation Method} (New York: Academic)
\item Hirota R 1972 { Exact solution of the sine-Gordon equation for multiple collisions of solitons}
    {\it J. Phys. Soc. Japan} {\bf 33} 1459
\item Caudrey RJ, Gibbon JD, Eilbeck JC and Bullough RK 1973 { Exact multisoliton solutions of the self-induced
transparency and sine-Gordon equation} {\it Phys. Rev. Lett.}  {\bf 30} 237
\item Ablowitz MJ, Kaup DJ, Newell AC and Segur H 1973 { Method for solving the sine-Gordon equation} 
{\it Phys. Rev. Lett.}  {\bf 30} 1262
\item Takhtadzhyan LA 1974 { Exact theory of propagation of ultrashort optical pulses in two-level media}
{\it Soviet Phys. JETP} {\bf 39} 228
\item Matsuno Y 2007 { Multiloop soliton and multibreather solutions of the short pulse model equation}
  {\it J. Phys. Soc. Japan} {\bf 76} 084003
\item Sh\"affer T and Wayne CE 2004 { Propagation of ultra-short optical pulses in cubic nonlinear media}
{\it Phys. D} {\bf 196} 90
\item  Tsitsas NL, Horikis TP, Shen Y, Kevrekidis PG, Whitaker N and Frantzeskakis DJ  2010
{ Short pulse equations and localized structures in frequency band gaps of nonlinear metamaterials}
{\it Phys. Lett. A} {\bf 374} 1384
\item Matsuno Y 2008 { Periodic solutions of the short pulse model equation} {\it J. Math. Phys.} {\bf 49} 073508
\item Matsuno Y 2009 { Soliton and periodic solutions of the short pulse model equation} 
in {\it Handbook of Solitons: Research, Technology and Applications} ed SP Lang and SH Bedore (New York: Nova) Chapter 15
\item Pogrebkov AK 1981 { Singular solitons: an example of a sinh-Gordon equation} {\it Lett. Math. Phys.} {\bf 5} 277
\item Lamb, Jr GL 1970 { Higher conservation laws in ultrashort optical  pulse propagation} {\it Phys. Lett. A} {\bf 32} 251
 \item Sanuki H and Konno K 1974 { Conservation laws of sine-Gordon equation}  {\it Phys. Lett. A} {\bf 48} 221

\end{enumerate}

\newpage
\leftline{\bf Figure captions} \par
\begin{itemize}
\item[{\bf Figure 1.}] The profile of a  kink $u$ (solid line) and corresponding profile of $v\equiv u_X$ (broken line). 
The parameter $p_1$ is set to $0.4$ and the parameter $y_0$ is chosen such that the center position of $u_X$ is at $X=0$.

\item[{\bf Figure 2.}]  The profile of a two-soliton solution $v\equiv u_x$
 for three different times, a: $t=0$,\ b: $t=2$,\ c: $t=4$.
The parameters are chosen as $p_1=0.2,\ p_2=0.5,\ \xi_{10}=-8,\ \xi_{20}=0$.

\item[{\bf Figure 3.}]  The profile of a breather solution  
 for three different times, a: $t=0$, \ b: $t=5$,\ c: $t=10$.
The parameters are chosen as $p_1=0.3+0.5\,{\rm i},\ p_2=p_1^*=0.3-0.5\,{\rm i},\ \xi_{10}=\xi_{20}^*=0$.

\item[{\bf Figure 4.}]  The profile of  $v\equiv u_x$  
 for three different times which represents the interaction between a soliton and a breather, a: $t=0$, \ b: $t=15$, c: $t=30$.
The parameters are chosen as $p_1=0.2+0.4\,{\rm i},\ p_2=p_1^*=0.2-0.4\,{\rm i}, \ p_3=0.3,\ \xi_{10}=\xi_{20}=0,\ \xi_{30}=-30$.

\item[{\bf Figure 5.}]  The profile of a breather-breather solution $u$  
 for three different times, a: $t=0$, \ b: $t=15$, c: $t=30$.
The parameters are chosen as $p_1=0.1+0.2\,{\rm i}, p_2=p_1^*=0.1-0.2\,{\rm i},  p_3=0.15+0.3\,{\rm i}, 
\ p_4=p_3^*=0.15-0.3\,{\rm i}, \ \xi_{10}=\xi_{20}^*=-15,
\ \xi_{30}=\xi_{40}^*=0$. 
\end{itemize}

\newpage
\begin{center}
\includegraphics[width=10cm]{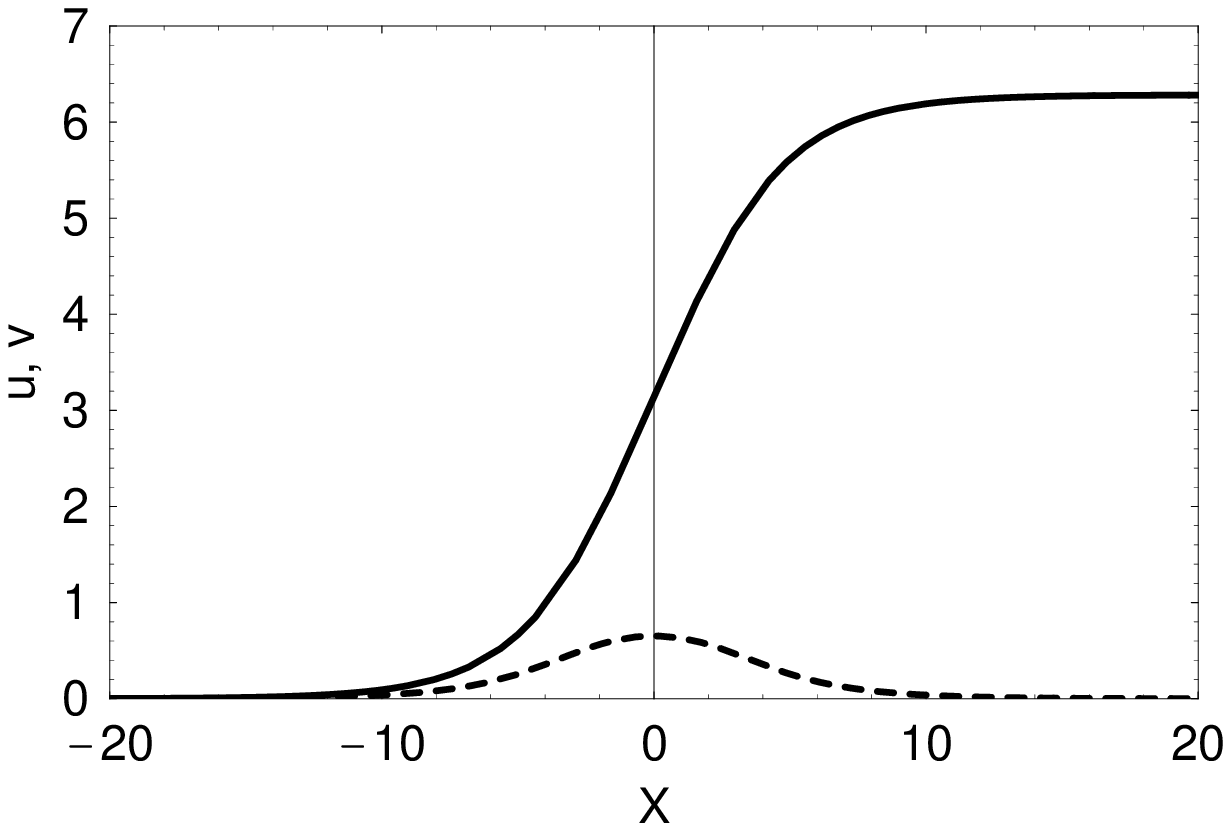}
\end{center}
\centerline{\bf Figure 1}

\newpage
\begin{center}
\includegraphics[width=10cm]{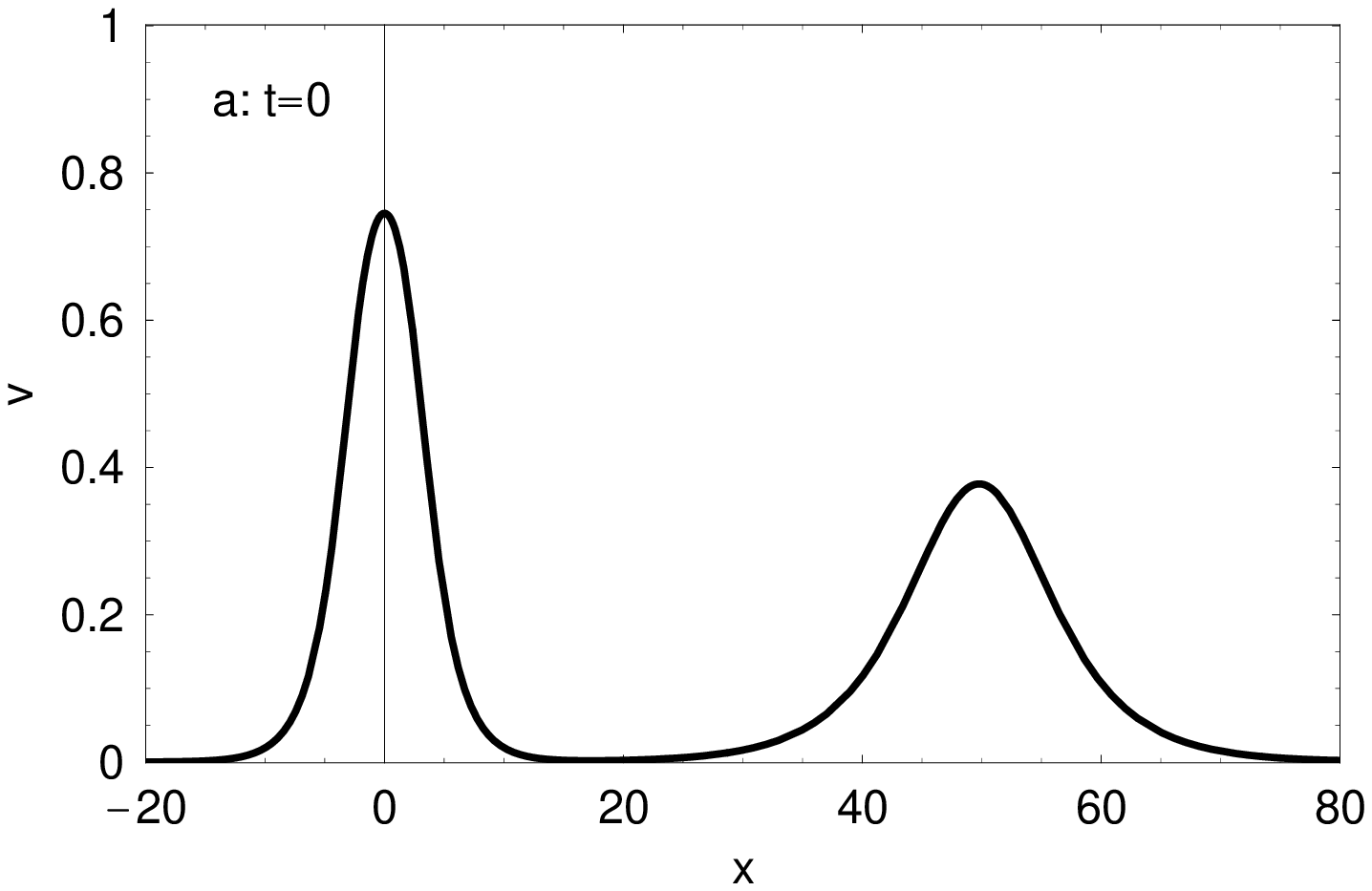}
\end{center}
\bigskip
\begin{center}
\includegraphics[width=10cm]{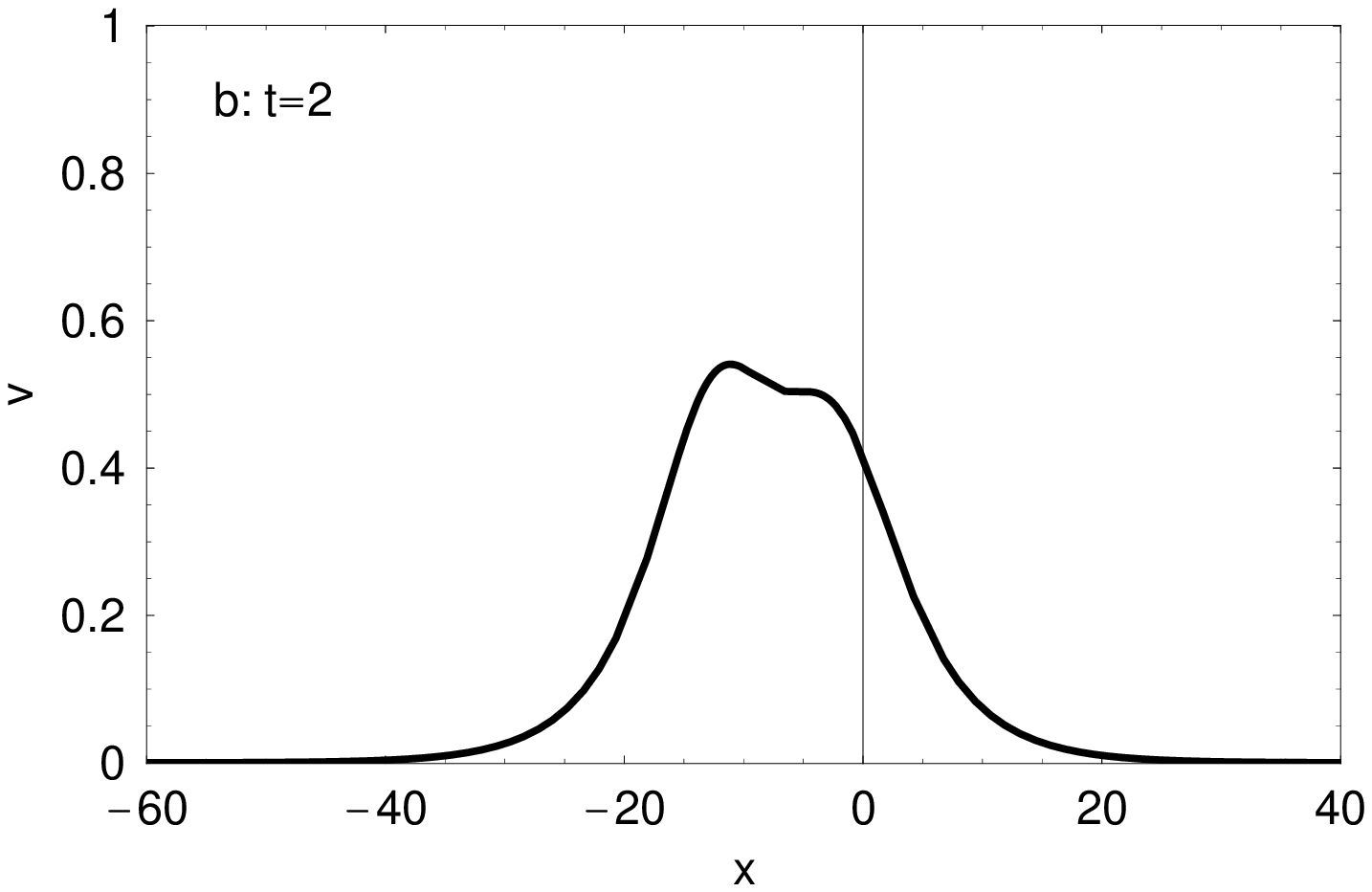}
\end{center}
\bigskip
\begin{center}
\includegraphics[width=10cm]{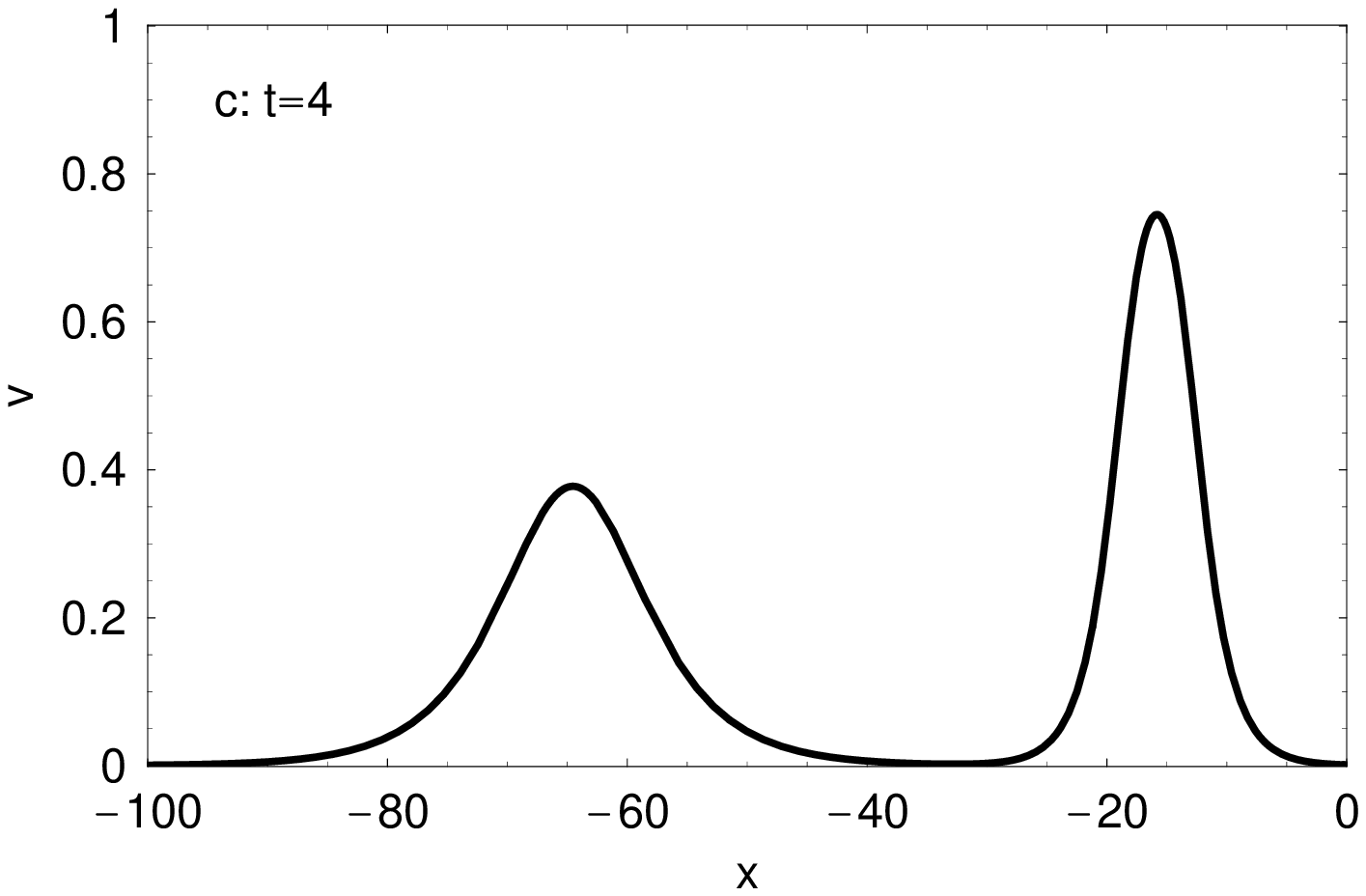}
\end{center}
\centerline{\bf Figure 2 a-c}

\newpage
\begin{center}
\includegraphics[width=10cm]{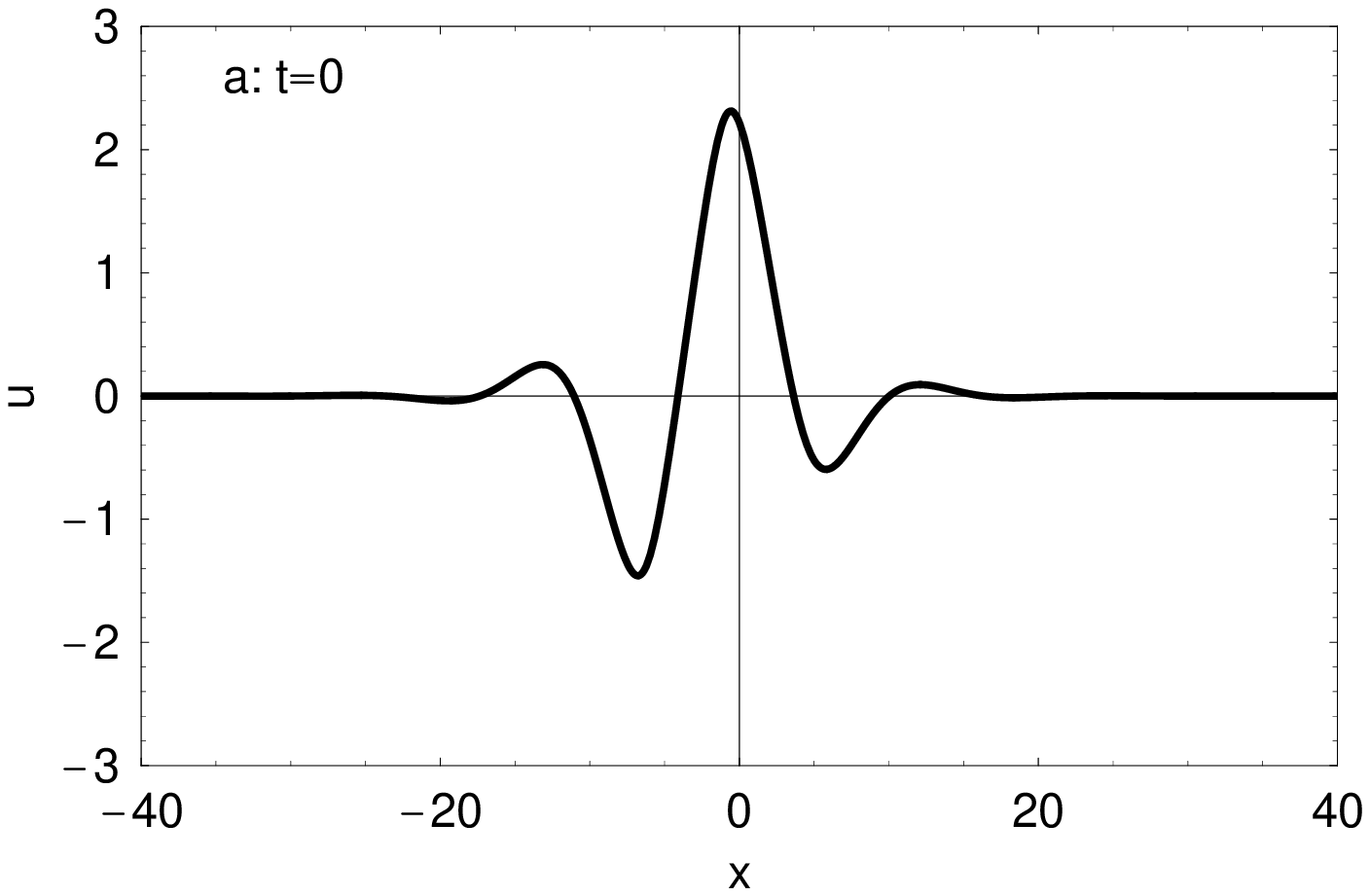}
\end{center}
\bigskip
\begin{center}
\includegraphics[width=10cm]{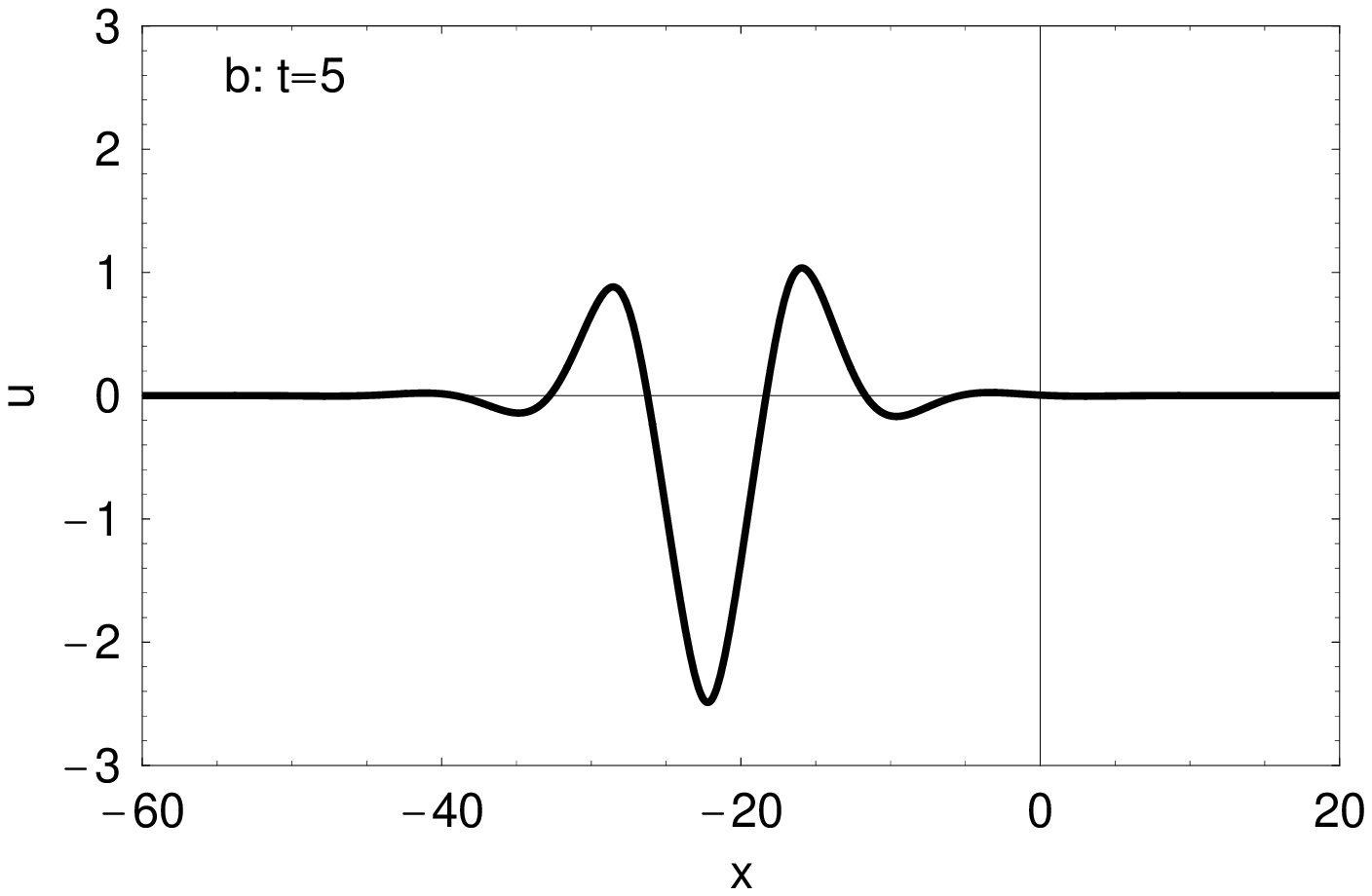}
\end{center}
\bigskip
\begin{center}
\includegraphics[width=10cm]{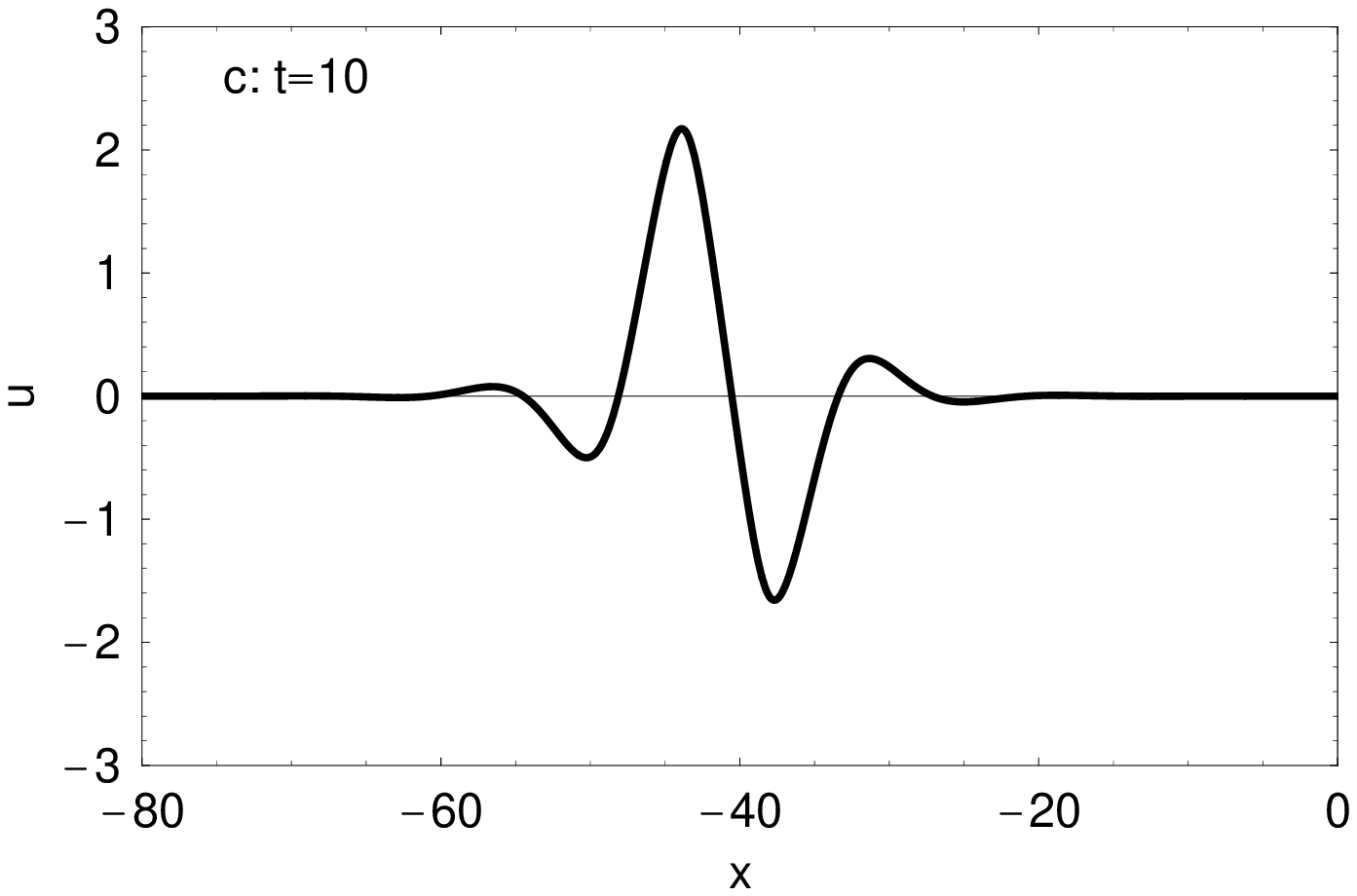}
\end{center}
\centerline{\bf Figure 3 a-c}

\newpage
\begin{center}
\includegraphics[width=10cm]{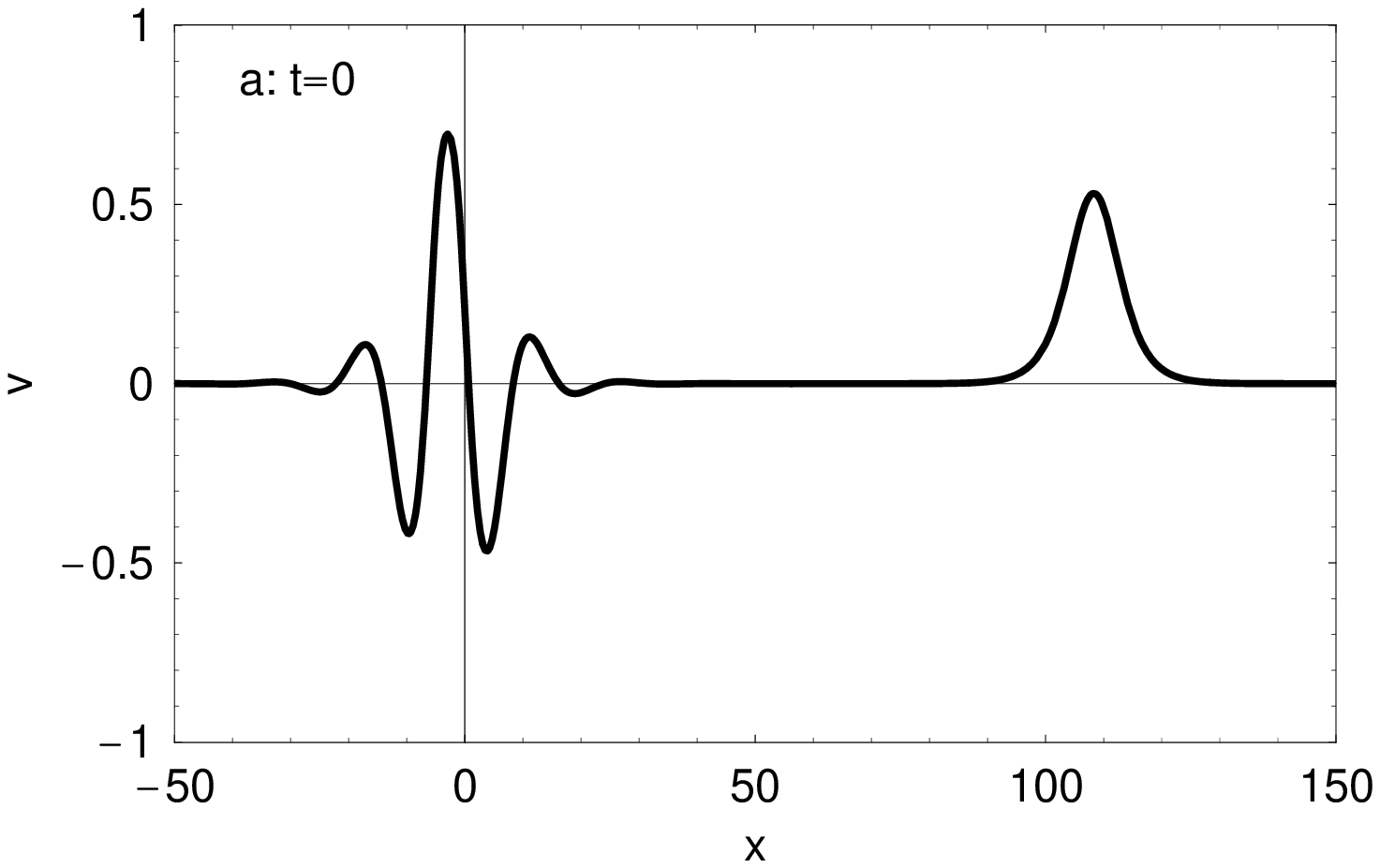}
\end{center}
\bigskip
\begin{center}
\includegraphics[width=10cm]{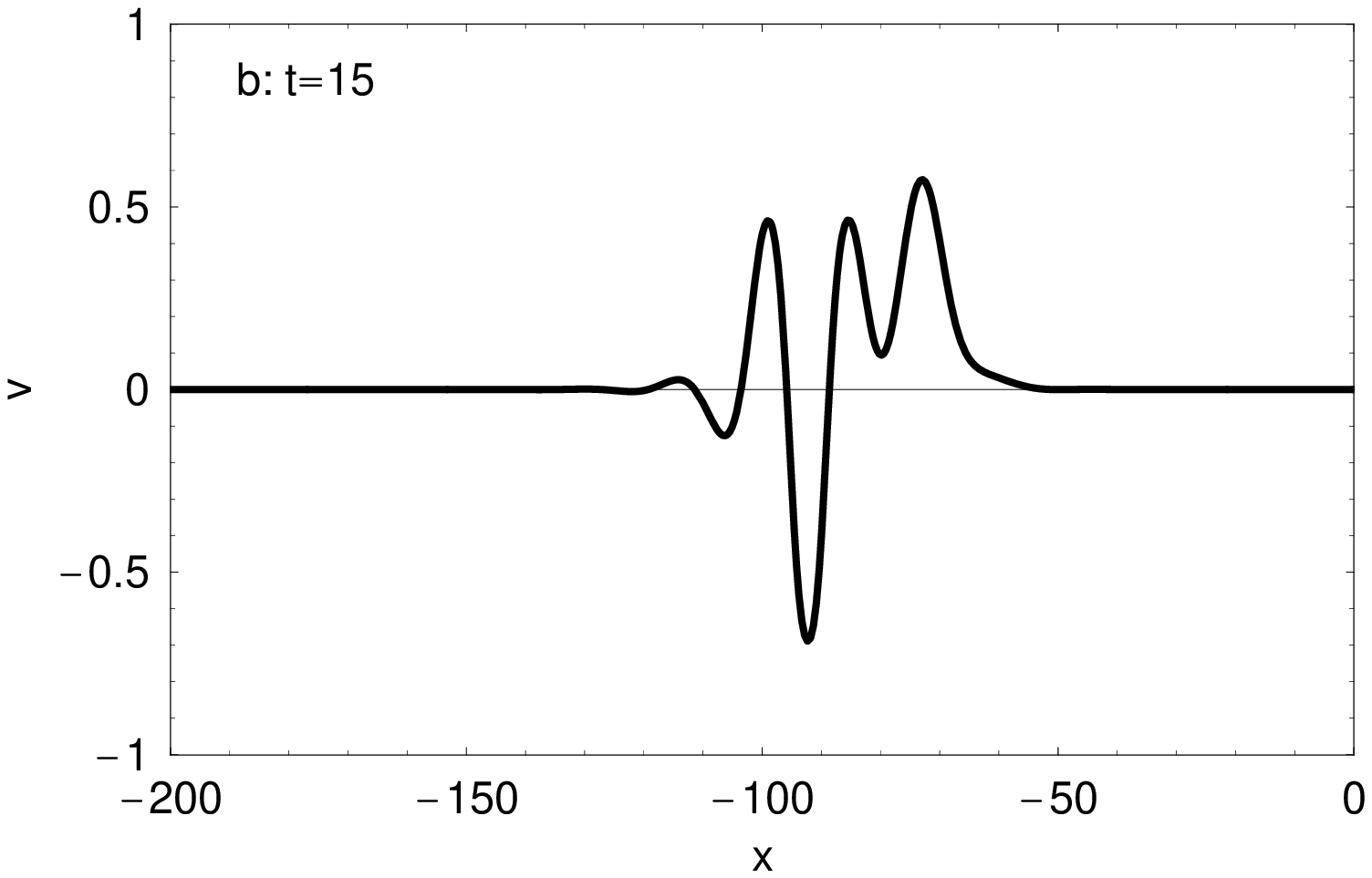}
\end{center}
\bigskip
\begin{center}
\includegraphics[width=10cm]{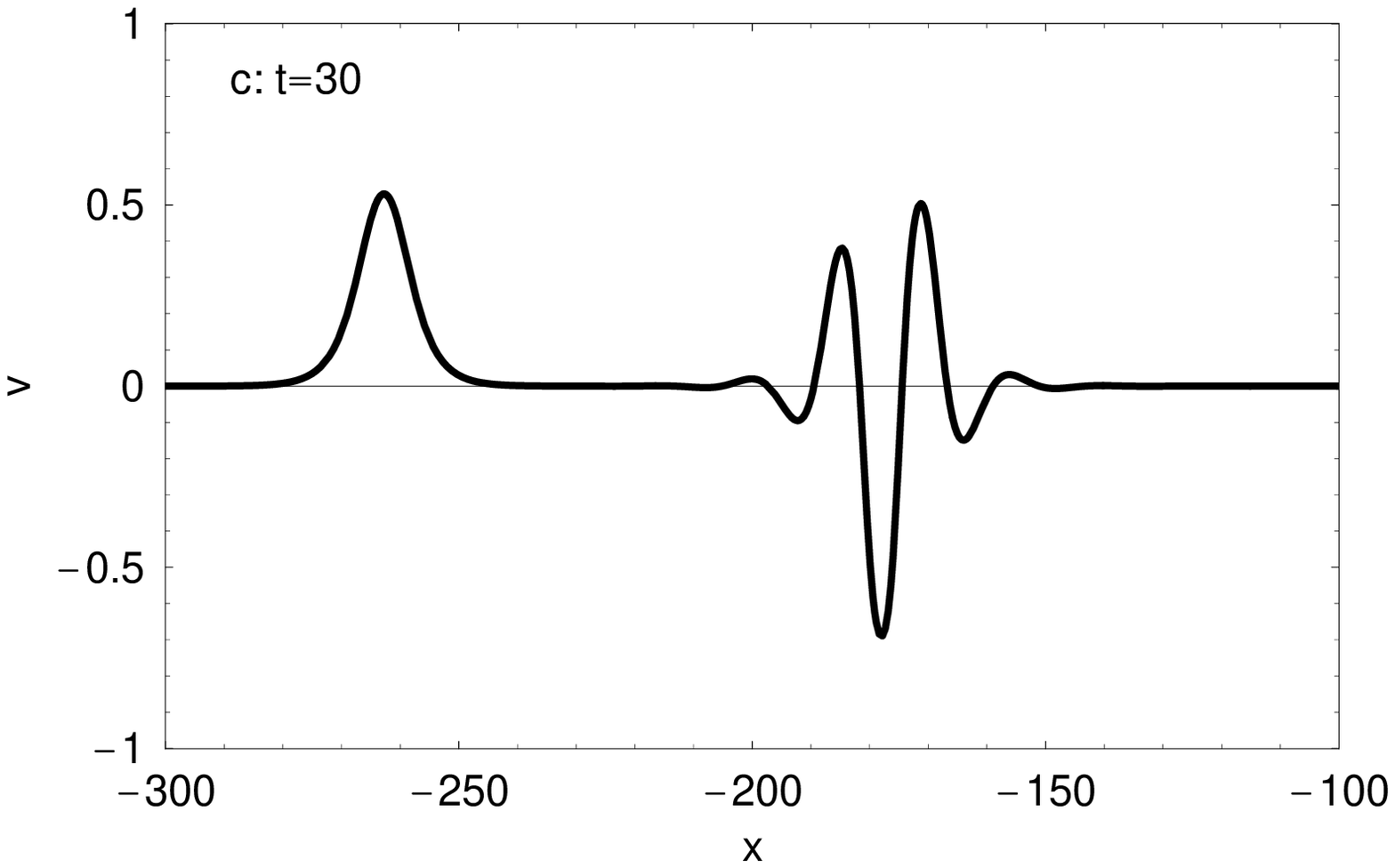}
\end{center}
\centerline{\bf Figure 4 a-c}

\newpage
\begin{center}
\includegraphics[width=10cm]{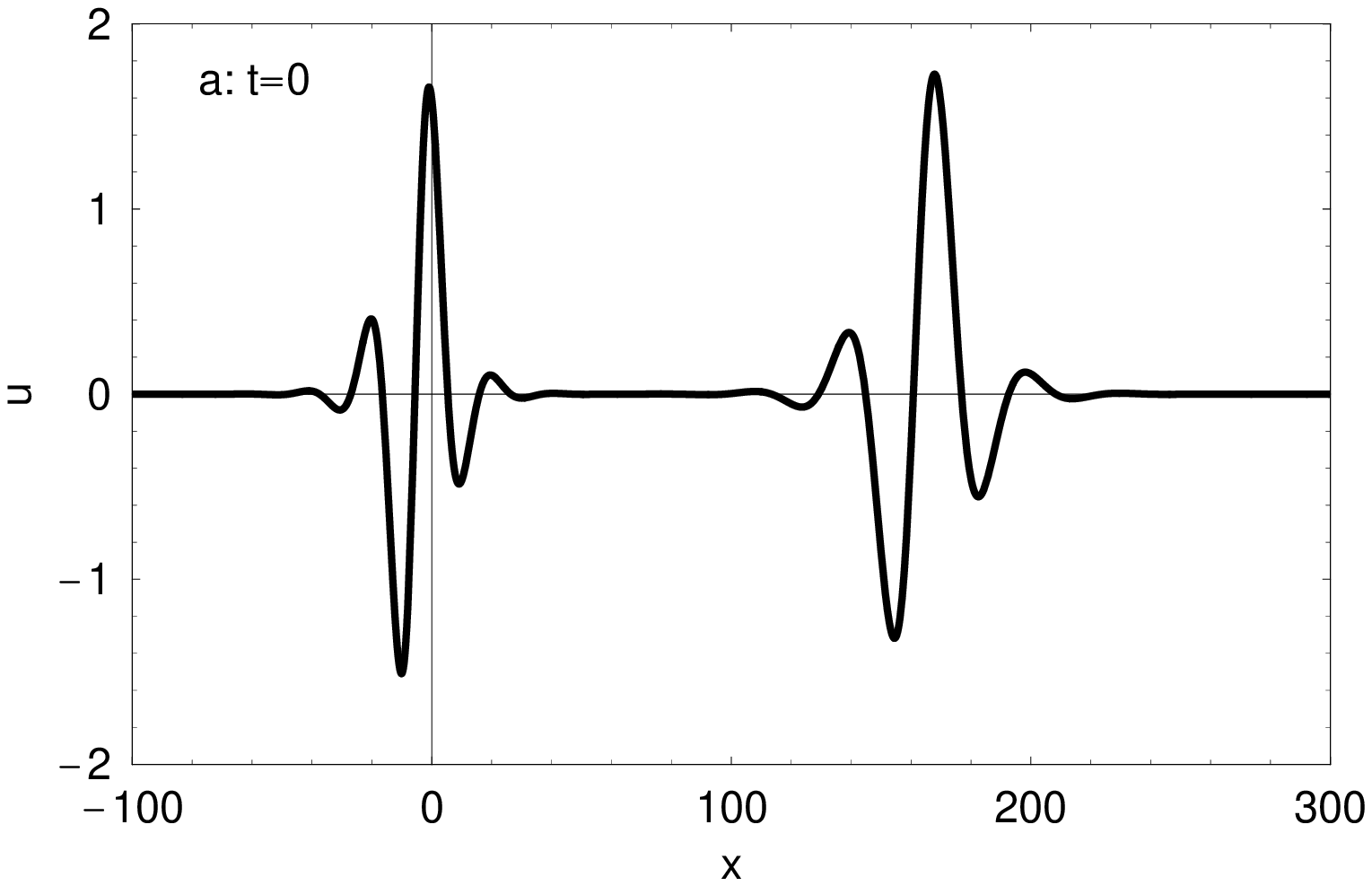}
\end{center}
\bigskip
\begin{center}
\includegraphics[width=10cm]{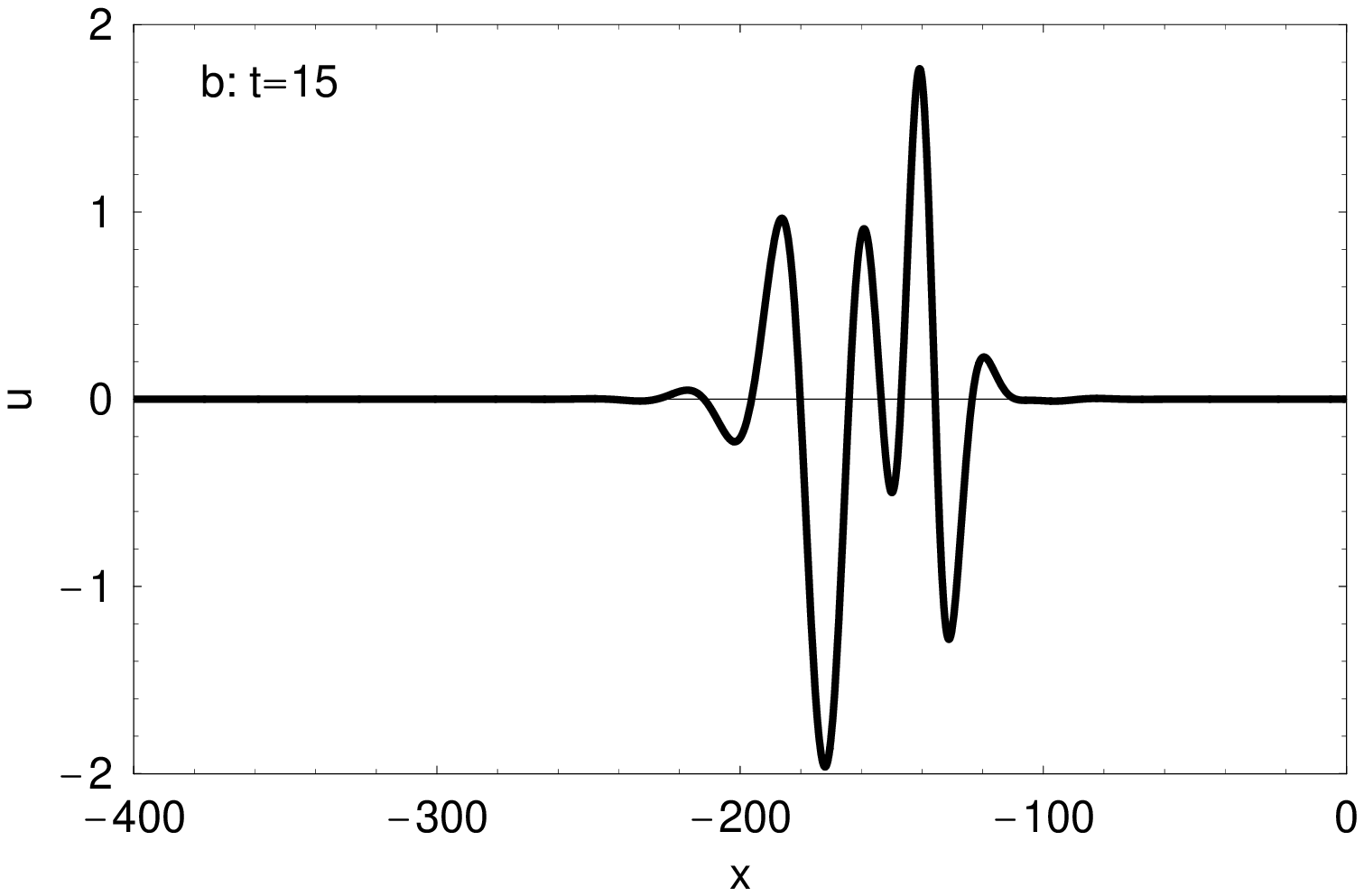}
\end{center}
\bigskip
\begin{center}
\includegraphics[width=10cm]{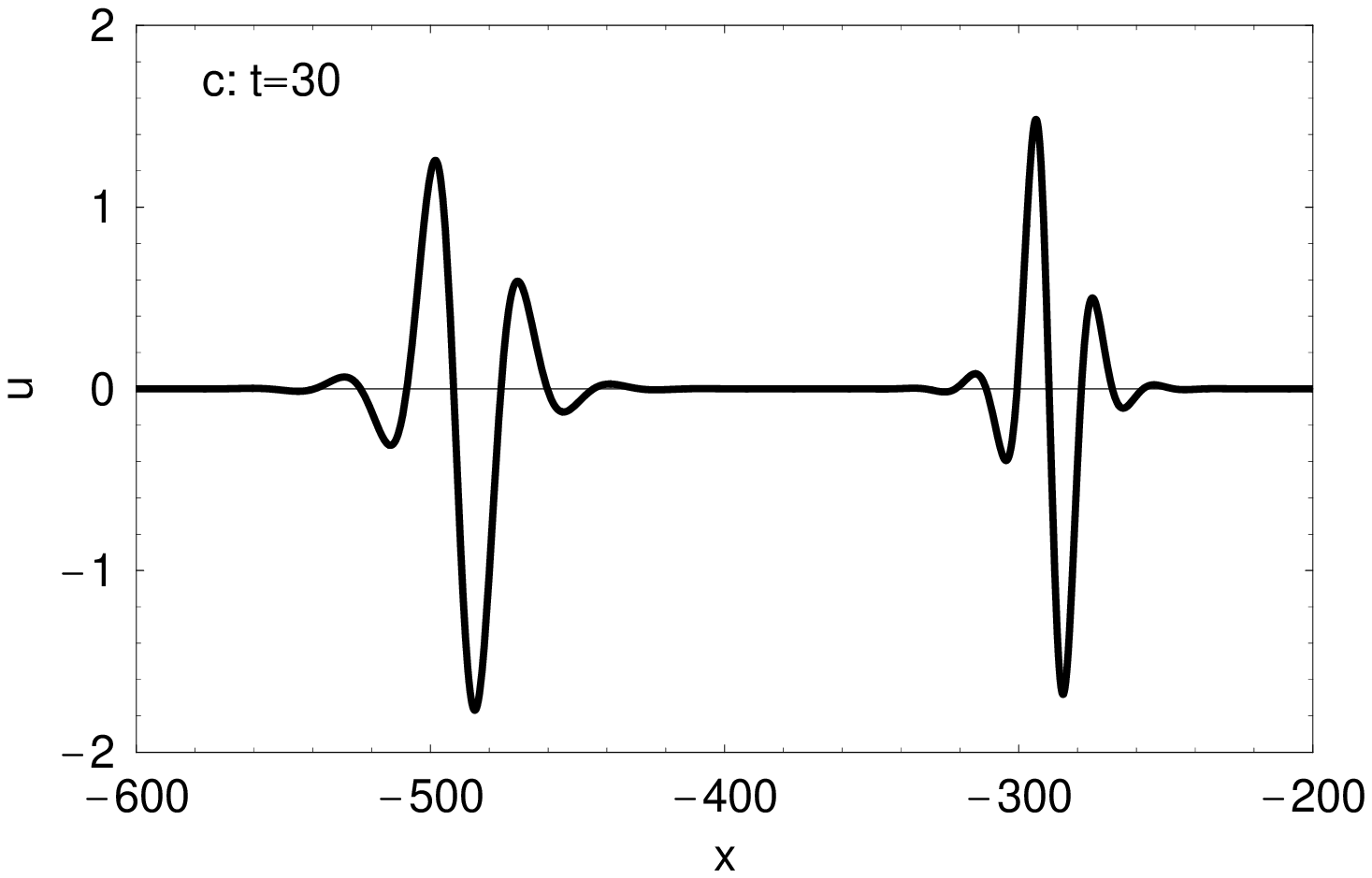}
\end{center}
\centerline{\bf Figure 5 a-c}

\end{document}